\documentclass[iop]{emulateapj}

\usepackage{graphicx}
\usepackage{epsfig}
\usepackage{color}
\usepackage{journals}
\usepackage{amsmath,amssymb,latexsym}

\allowdisplaybreaks

\shorttitle{Density Independent Formulation of SPH}
\shortauthors{Saitoh \& Makino}

\begin{document}

\title{A Density Independent Formulation of Smoothed Particle Hydrodynamics}

\author{Takayuki \textsc{R.Saitoh}\altaffilmark{1}
        \& Junichiro \textsc{Makino}\altaffilmark{1}
}
\altaffiltext{1}{Earth-Life Science Institute, Tokyo Institute of
Technology, 2--12--1, Ookayama, Meguro, Tokyo, 152-8551, Japan}
\email{saitoh@geo.titech.ac.jp}

\begin{abstract}
The standard formulation of the smoothed particle hydrodynamics (SPH) assumes
that the local density distribution is differentiable.  This assumption is used
to derive the spatial derivatives of other quantities. However, this assumption
breaks down at the contact discontinuity.  At the contact discontinuity, the
density of the low-density side is overestimated while that of the high-density
side is underestimated.  As a result, the pressure of the low (high) density
side is over (under) estimated.  Thus, unphysical repulsive force appears at the
contact discontinuity, resulting in the effective surface tension.  This tension
suppresses fluid instabilities.  In this paper, we present a new formulation of
SPH, which does not require the differentiability of density.  Instead of the
mass density, we adopt the internal energy density (pressure), and its arbitrary
function, which are smoothed quantities at the contact discontinuity, as the
volume element used for the kernel integration.  We call this new formulation
density independent SPH (DISPH). It handles the contact discontinuity without
numerical problems. The results of standard tests such as the shock tube,
Kelvin-Helmholtz and Rayleigh-Taylor instabilities, point like explosion, and
blob tests are all very favorable to DISPH. We conclude that DISPH solved most
of known difficulties of the standard SPH, without introducing additional
numerical diffusion or breaking the exact force symmetry or energy conservation.
Our new SPH includes the formulation proposed by \citet{RitchieThomas2001} as a
special case. Our formulation can be extended to handle a non-ideal gas easily.
\end{abstract}


\section{Introduction} \label{sec:intro}

Smoothed particle hydrodynamics (SPH) is a Lagrangian scheme to solve the
evolution of fluid using particles. It was originally introduced by
\citet{Lucy1977} and \citet{GingoldMonaghan1977} and has been widely used in the
field of the computational astrophysics \citep[][]{Monaghan1992, Monaghan2005,
Rosswog2009, Springel2010Review}.  It is becoming popular in hydrodynamical
simulations in engineering \citep[e.g.,][]{LiuLiu2003}.

Recently, \citet{Agertz+2007} reported the results of comparison of SPH and
Euler schemes (grid methods). Their main finding is that SPH suppresses the
Kelvin-Helmholtz instability. This has been pointed out earlier by
\citet{Okamoto+2003}.  The reason of this problem is that in the standard SPH
the smoothed density is used to obtain other physical quantities.  The estimated
density of particles near the contact discontinuity has $\mathcal O(1)$ error,
irrespective of the numerical resolution.  This large error causes similarly
large error in the pressure (see \S \ref{sec:SSPH}).  \citet{Agertz+2007} noted
that there were {\it fundamental differences} between SPH and grid methods.

There have been several proposals to improve SPH so that it can deal with the
contact discontinuity. \citet{Price2008} discussed the artificial thermal
conductivity which was originally introduced by \citet{Monaghan1997}. The
motivation of the use of the artificial conductivity is that every physical
quantity should be {\it smooth} in the standard SPH.  The artificial
conductivity eliminates the discontinuity in the thermal energy.  Since both the
density and energy (entropy) near the contact discontinuity become smooth, the
pressure across the contact discontinuity becomes smooth.  Thus, the
Kelvin-Helmholtz instability takes place.  At the first sight, this artificial
conductivity looks similar to the artificial viscosity which is necessary to
capture shocks in SPH.  However, there are two fundamental differences. First,
the artificial viscosity is used to generate the physical dissipation associated
with the shock, while the artificial conductivity adds physically non-existent
dissipation.  One needs to fine-tune the conductivity coefficient to prevent
unnecessary smoothing. This means that the conductivity must be nonlinear.
Second, if there is a jump in the chemical composition, thermal conductivity is
not enough.  However, whether the use of artificial chemical diffusion is
justified or not is an open question.  He also showed that, when the Ritchie \&
Thomas formulation \citep{RitchieThomas2001} was used, the Kelvin-Helmholtz
instability grew but the growth of the instability was insufficient.
\citet{Read+2010} suggested that the Kelvin-Helmholtz instability took place
when a higher order kernel with a large enough number of neighbor particles
and the momentum equation of the \citet{RitchieThomas2001} were used.
\citet{Abel2011} used the relative pressure, which was first proposed by
\citet{Morris1996}, instead of the absolute values of pressures in the equation
of motion.  This formulation improves the treatment of the Kelvin-Helmholtz
instability, but breaks the Newton's third law.  \citet{Garcia-Senz+2012}
considered the use of the integral form of the first derivative, which also
improved the treatment of hydrodynamical instabilities.

In this paper, we describe a new formulation of SPH which does not use the
smoothed mass density for the volume element. Instead, we use an arbitrary
function of the internal energy density (pressure) for the volume element to
obtain other quantities and their spatial derivatives.  The reason why we adopt
the energy density instead of the mass density is that it is the fundamental
quantity of the hydrodynamics. We call this new formulation density independent
SPH (DISPH).  In DISPH, the pressure is calculated without using the mass
density.  Thus, unphysical jumps of pressure at the contact discontinuity
disappear.  The special case that the volume element defined by the internal
energy density or pressure leads to the equation of motion proposed by
\citet{RitchieThomas2001}.  Our formulation can be used to derive the SPH
equation for an arbitrary quantity, while how we can apply the Ritchie \& Thomas
formulation to equations other than energy equation and equation of motion is
not clear.  Results of various tests indicate that our formulation is highly
advantageous.

The structure of this paper is as follows.  In \S \ref{sec:SSPH}, we analyze the
problem of standard SPH at discontinuities. The derivation of DISPH is described
in \S \ref{sec:DISPH}. We then generalized DISPH adopting an arbitrary function
of pressure in \S \ref{sec:GDISPH}.  The comparison of the results of test
calculations with DISPH and standard formulation of SPH are shown in \S
\ref{sec:results}. Summary and discussion are presented in \S \ref{sec:summary}.

\section{Standard SPH and Its Difficulty around Discontinuities} \label{sec:SSPH}

In SPH, the fluid is expressed by discrete particles and physical quantities are
approximated by kernel interpolation. In the standard formulation of SPH, the
local density is first calculated, and then the rests of necessary physical
quantities, such as the pressure gradient and the time derivative of the
internal energy, are calculated.  Thus, the accuracy of the solution depends on
the accuracy of the density estimate. In this section, we reexamine the
derivation of the equation of motion in SPH to understand its problem.

A physical quantity $f$ at position $\boldsymbol r$ can be expressed as follows:
\begin{equation}
f({\boldsymbol r}) = \int f(\boldsymbol r') \delta(|\boldsymbol r - \boldsymbol r'|) d \boldsymbol r'. \label {eq:f}
\end{equation}
A smoothed value of $f$ at position $\boldsymbol r$, $\langle f
\rangle(\boldsymbol r)$, is given by the convolution of $f$ and a kernel
function
$W(\boldsymbol r - \boldsymbol r',h)$:
\begin{equation}
\langle f \rangle (\boldsymbol r) = \int f(\boldsymbol r') W(|\boldsymbol r -
\boldsymbol r'|,h) d \boldsymbol r', \label{eq:smoothed_f}
\end{equation}
where $h$ is the size of the kernel function and corresponds to the spatial
resolution. This smoothing is the base of SPH.  Here, the kernel function must
satisfy the following three conditions: (1) it becomes the delta function in the
limit of $h \rightarrow 0$, (2) it is normalized as unity, and (3) it is a
function with compact support.  A cubic spline function is most widely used as
the kernel function:
\begin{equation}
W(|\boldsymbol r - \boldsymbol r'|,h) = \frac{\sigma}{h^D}
\begin{cases}
\left (1-\frac{3}{2}s^2+\frac{3}{4}s^3 \right ) & 0\le s <1, \\
\frac{1}{4} (2-s)^3 & 1 \le s <2, \\
0 & 2 \le s,  \label{eq:kernel:spline}
\end{cases}
\end{equation}
where $s = |\boldsymbol r - \boldsymbol r'|/h$, $D$ is the dimension, and the
normalized factors $\sigma$ in one, two, and three dimensions are $2/3$,
$10/7\pi$, and $1/\pi$, respectively. We first derive the equations of motion
and energy with the constant kernel size, and then we generalized them to the
individual kernel size.

The first derivative of the smoothed $f$ is given by
\begin{equation}
\langle \nabla f \rangle (\boldsymbol r) = \int \nabla f(\boldsymbol r') 
W(|\boldsymbol r - \boldsymbol r'|,h) d \boldsymbol r'. \label{eq:f:derivation}
\end{equation}
By making use of the partial integral and the fact that the kernel function has
compact support, Eq. \eqref{eq:f:derivation} becomes
\begin{equation}
\langle \nabla f \rangle (\boldsymbol r) = \int f(\boldsymbol r') 
\nabla W(|\boldsymbol r - \boldsymbol r'|,h) d \boldsymbol r'. \label{eq:smoothed_f:derivation}
\end{equation}

We need to discretize Eq. \eqref{eq:smoothed_f} to evaluate the physical
quantities at positions of particles. To convert integral into summation, a
volume element $d \boldsymbol r'$ is replaced by the discrete volume element
$\Delta V_j = m_j/\rho_j$, where $m_j$ and $\rho_j$ are the mass and density of
the particle $j$. In addition, positions of particles $i$ and $j$ are expressed
by $\boldsymbol r_i$ and $\boldsymbol r_j$ and $f(\boldsymbol r')$ is replaced
by $f_j$.  Thus, the value of $f$ at the position of particle $i$ is
\begin{equation}
f_i = \sum_j m_j \frac{f_j}{\rho_j} W_{ij}(h), \label{eq:smoothed_f:sum}
\end{equation}
where $f_i \equiv \langle f \rangle (\boldsymbol r_i)$ and $W_{ij} =
W(|\boldsymbol r_i - \boldsymbol r_j|,h)$.
Hereafter, we call the SPH formulation with this type of discretization as the
{\it standard SPH}.  At this point, we do not know $\rho_j$. By substituting
$\rho$ into $f$, we obtain
\begin{equation}
\rho_i = \sum_j m_j W_{ij}(h), \label{eq:ssph:density}
\end{equation}
where $\rho_i \equiv \langle \rho \rangle (\boldsymbol r_i)$ is the smoothed
density at the position of particle $i$. Note that the right-hand side of Eq.
\eqref{eq:ssph:density} includes no unknown quantities.  Thus, densities should be
calculated first in the standard SPH.

The equation of motion is
\begin{equation}
\frac{d^2 \boldsymbol r}{d t^2} = -\frac{\nabla P}{\rho}, \label{eq:euler}
\end{equation}
where $t$ is time and $P$ is pressure. The SPH approximation of Eq.
\eqref{eq:euler} is given by 
\begin{equation} 
\frac{d^2 \boldsymbol r_i}{d t^2} = -\sum_j m_j \left ( \frac{P_i}{\rho_i^2}
+ \frac{P_j}{\rho_j^2} \right ) \nabla W_{ij}(h). \label{eq:ssph:euler}
\end{equation}
This form satisfies the Newton's third law. We used the following relation to
obtain Eq. \eqref{eq:ssph:euler}:
\begin{equation}
\frac{\nabla P}{\rho} = \nabla \left ( \frac{P}{\rho} \right )  -
\frac{P}{\rho^2} \nabla \rho. \label{eq:ssph:euler:symmetrized}
\end{equation}
In order for Eq. \eqref{eq:ssph:euler} to be meaningful, $\rho$ must be
differentiable, since its derivative is used in Eq.
\eqref{eq:ssph:euler:symmetrized}.

Finally, we derive the energy equation in the standard SPH. The energy equation
is
\begin{equation}
\frac{d u}{dt} = -\frac{P}{\rho} \nabla \cdot \boldsymbol v,
\end{equation}
where $u$ is the internal energy and $\boldsymbol v$ is the velocity.  To obtain
the SPH formulation of the energy equation, we need the SPH expression of
$\nabla \cdot \boldsymbol v$.  We use
\begin{equation}
\nabla (\rho \boldsymbol v) = 
\nabla \rho \boldsymbol v + \rho \nabla \cdot \boldsymbol v. \label{eq:ssph:massflux}
\end{equation}
The SPH formulation of $\nabla \cdot \boldsymbol v$ is given by
\begin{align}
\rho_i \nabla \cdot \boldsymbol v_i &= 
\sum_j m_j \boldsymbol v_j \cdot \nabla W_{ij}(h) 
        -\boldsymbol v_i \cdot \sum_j m_j \nabla W_{ij}(h)  \notag\\
&= - \sum_j m_j \boldsymbol v_{ij} \cdot \nabla W_{ij}(h),
\label{eq:ssph:divv}
\end{align}
where $\boldsymbol v_{ij} = \boldsymbol v_{i} - \boldsymbol v_j$.
Therefore, the energy equation in the standard SPH is 
\begin{equation}
\frac{du_i}{dt} = \sum m_j \frac{P_i}{\rho_i^2} \boldsymbol v_{ij} 
\cdot \nabla W_{ij}(h). \label{eq:ssph:energy}
\end{equation}
Equations \eqref{eq:ssph:density}, \eqref{eq:ssph:euler}, and \eqref{eq:ssph:energy}
close with the equation of state (EOS),
\begin{equation}
P = (\gamma-1) \rho u, \label{eq:EOS}
\end{equation}
where $\gamma$ is the specific heat ratio. There is no need to solve the
continuity equation in SPH since it is satisfied automatically.

When we use the variable and individual kernel size, above equations should be
modified slightly.  Here, we adopt a simple and traditional way.  First, the
density evaluation equation is rewritten as 
\begin{equation}
\rho_i = \sum_j m_j W_{ij}(h_i). \label{eq:ssph:density:ind_h}
\end{equation}
This is the same as so-called gather interpretation of the summation
\citep{HernquistKatz1989}.  When a variable kernel size is employed, an
iterative approach is used to determine both $\rho$ and $h$ imposing a
condition, for instance, that the number of neighbor particles is kept in a
fixed range.  In equations of motion and energy, the gather-and-scatter
interpretation is used \citep{HernquistKatz1989}. Thus, Eqs.
\eqref{eq:ssph:euler} and \eqref{eq:ssph:energy} become 
\begin{equation}
\frac{d^2 \boldsymbol r_i}{d t^2} = -\sum_j m_j \left ( \frac{P_i}{\rho_i^2}
+ \frac{P_j}{\rho_j^2} \right ) \nabla \tilde{W}_{ij},
\label{eq:ssph:euler:ind_h}
\end{equation}
and 
\begin{equation}
\frac{du_i}{dt} = \sum_j m_j \frac{P_i}{\rho_i^2} \boldsymbol v_{ij} 
\cdot \nabla \tilde{W}_{ij}, \label{eq:ssph:energy:ind_h}
\end{equation}
where $\nabla W_{ij}(h_i)$ is replaced by $\nabla \tilde{W}_{ij} =
0.5 [\nabla W_{ij}(h_i)+ \nabla W_{ij}(h_j)]$ so that the equation of motion
can satisfy the Newton's third law. It is also possible to use $\nabla \tilde{W}
_{ij} = \nabla W_{ij}[0.5(h_i+h_j)]$. We adopt the first form
throughout this paper.

The Lagrangian formulation \citep{SpringelHernquist2002} provides the derivative
of the kernel size.  We will show the comparison results in \S
\ref{sec:explosion}.

In the derivation of the standard SPH discretization, the differentiability of
$\rho$ is used for both the equation of motion and the energy equation.
However, $\rho$ is discontinuous at the contact discontinuity. In the following,
we illustrate the consequence of the discontinuity of the density.

\begin{figure*}[htb]
\begin{center}
\epsscale{0.9}
\plotone{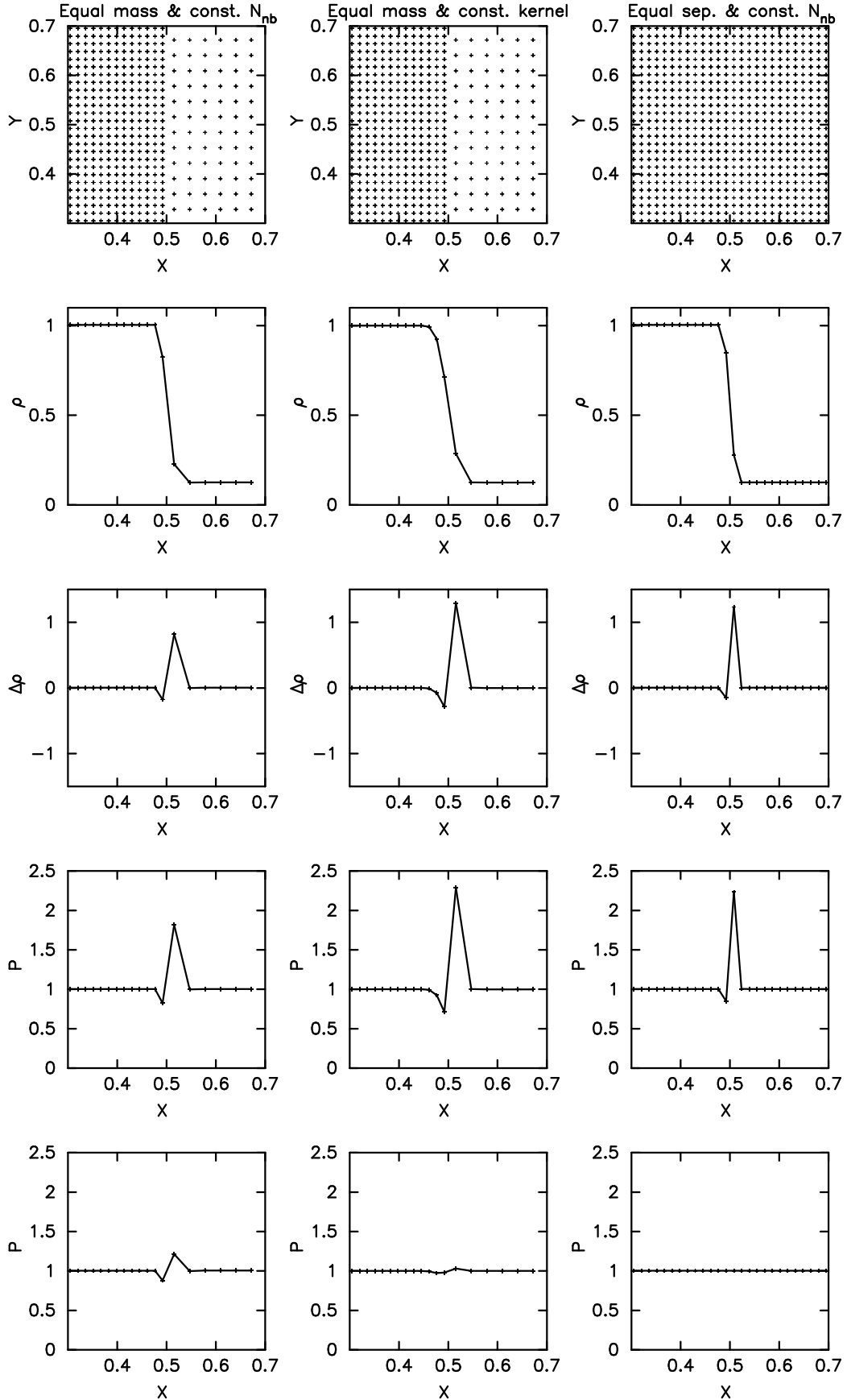}
\caption{Density and pressure fields evaluated with the standard SPH and our 
SPH around the contact discontinuity with the density ratio of $1:8$.  Equal
mass particles are used for the first and second configurations (the left and
middle columns). The positions of the less dense region is determined by taking
the center of mass of the eight particles in the cube of the particle
separation.  The equal separation is used in the last configuration (the right
column).  In this configuration, the mass of particles in the less dense region
is 1/8 of that of particles in the dense region.  For the left and right
columns, the constant neighbor number, $32 \pm 2$, is used.  In the middle
column, a constant kernel size of 0.03125 is used.  The top row shows the
distribution of particles projected on the $x-y$ plane.  The second row shows
the density of each SPH particle evaluated with Eq.
\eqref{eq:ssph:density:ind_h}. The third row shows the density contrast between
the evaluated density and true one.  The fourth row shows corresponding
pressure.  In the bottom row, the pressure of each particle calculated with
DISPH is shown.
\label{fig:pressure}
}
\end{center}
\end{figure*}

In Figure \ref{fig:pressure}, we show the values of density and pressure
around a contact discontinuity evaluated by the standard formulation of SPH.
Equation \eqref{eq:ssph:density:ind_h} is used and $P = (\gamma -1)\rho u$.
To set up this contact discontinuity, we place particles on a regular grid in three
dimensions and set $\rho = 1$ for $x<0.5$ and $\rho = 0.125$ for $x>0.5$. We
used equal-mass particles in the first two configurations.  In these two setups,
positions of particles in the less dense region is determined by taking the
center of mass of the eight particles in the cube of the particle separation.
In the last configuration, we adopted the equal separation for both regions,
which means that the mass of particles in the less dense region is 1/8 of that
of particles in the dense region.  The internal energy was set to $1.5$
($x<0.5$) and $12$ ($x>0.5$), and the specific heat ratio was $5/3$.  Velocities
of particles were set to zero.  The kernel size is determined to keep the
neighbor number, $N_{\rm nb}$, to the range $32 \pm 2$, in the first and the
last tests. In the second test a constant $h$ fixed to twice the particle
separation in the less dense region is used.

The top panels show the distribution of particles. The panels in the second row
show the SPH density.  Though the initial setup has the discontinuity at $x =
0.5$, it is smoothed by the kernel.  As a result, the SPH density of particles
next to the discontinuity has very large errors, as shown in the panels in the
3rd row. This large error in the density causes similarly large error in the
pressure (4th row).  The pressure of particles at the end of the low-density
region is grossly overestimated, while that at the end of the high-density
region is underestimated only modestly. This non-symmetric error in the pressure
is the origin of the repulsive force at the contact discontinuity, as has been
pointed out in previous studies \citep[e.g., ][]{RitchieThomas2001,
Okamoto+2003, Agertz+2007}.  This large error in the pressure also exists in
both of the constant kernel size case (the middle column) and the
equal-separation case (the right column). 

Consider the following density and pressure distribution:
\begin{equation}
\rho = 
\begin{cases}
\rho_1 & x\ge 0,\\
\rho_2 & x< 0,  \label{eq:discontinuity:rho}
\end{cases}
\end{equation}
and 
\begin{equation}
P = P_0.
\end{equation}
Obviously, we have 
\begin{equation}
\langle \rho \rangle (x) \rightarrow \frac{\rho_1+\rho_2}{2}, ~~~{\mathrm {for}}~x
\rightarrow 0, \label{eq:discontinuity:rho_mean}
\end{equation}
and therefore,
\begin{equation}
\lim_{x\rightarrow+0} \langle P \rangle (x) = \frac{\rho_1 + \rho_2}{2 \rho_1}
P_0,
\end{equation}
\begin{equation}
\lim_{x\rightarrow-0} \langle P \rangle (x) = \frac{\rho_1 + \rho_2}{2 \rho_2}
P_0.
\end{equation}
Thus, if $\rho_1 \ll \rho_2$, the error of the pressure can be arbitrarily
large. Note that the existence of this error does not imply the inconsistency of
SPH.  In this limit of $h \rightarrow 0$, the volume of the regime affected by
this error approaches to zero, which means the original differential equation is
restored almost everywhere. In other words, SPH satisfies the weak form of the
original equation. However, it means the convergence is slow and first order.

One might think that this error is caused by an inadequate initial thermal
energy (or density) distribution.  However, it is not the case. If we initialize
the internal energy (and/or density) of particles near the contact discontinuity
so that the pressure is smooth initially, this is no problem. However, as the
particles move, the change of the density can become sharpen, resulting in the
numerical problem described above.  We, thus, need continuous adjustment to
suppress the pressure error throughout the time integration.  Price's artificial
conductivity \citep{Price2008} provides such a continuous adjustment. Though the
artificial conductivity works beautifully in test calculations for the
Kelvin-Helmholtz instability, whether its use in actual astrophysical simulation
is justified or not is a bit questionable. First, in the case of the
discontinuity of chemical composition, not only the jump in the internal energy
but also that in the chemical composition should be smoothed but that is clearly
not adequate. Second, the artificial heat conduction can significantly enhance
the thermal relaxation of the system, which is again unwanted.

\section{A Density Independent Formulation of SPH} \label{sec:DISPH}

In \S \ref{sec:SSPH}, we have seen that the standard SPH breaks down at the
contact discontinuity because the continuity and differentiability of the
density is necessary to guarantee the convergence of SPH approximation.  The
basic reason for this problem is the use of $m_j/\rho_j$ for the volume element.
Thus, if we use something else as the volume element, we might be able to avoid
this difficulty altogether.  As shown in \S \ref{sec:GDISPH}, the formulation we
show in this section is the special case of the density independent formulation
of SPH.

\subsection{Concept} \label{sec:disph:basis}

Here, we propose an alternative formulation of SPH in which we discretize Eq.
\eqref{eq:smoothed_f} using the EOS of fluid, not the mass density.  The new
volume element of $j$-th particle is 
\begin{equation}
d \boldsymbol r' =
\Delta V_j = \frac{(\gamma -1 )m_j u_j}{P_j} = \frac{U_j}{q_j}, \label{eq:disph:volume_element}
\end{equation}
where $q_j \equiv \rho_j u_j$ and $U_j = m_j u_j$ are the energy density and the
internal energy of particle $j$, respectively.
Substituting Eq. \eqref{eq:disph:volume_element} into Eq. \eqref{eq:smoothed_f} and
using the gather summation, we obtain a new SPH approximation of smoothed $f$:
\begin{align}
f_i &=
\sum_j (\gamma-1) \frac{ m_j u_j f_j}{P_j} W_{ij}(h_i), \\
&= \sum_j (\gamma-1) \frac{U_j f_j}{P_j} W_{ij}(h_i). \label{eq:disph:smoothed_f:sum}
\end{align}
By substituting $f$ with the energy density, $q$, we have 
\begin{equation}
q_i = \sum_j U_j W_{ij}(h_i).  \label{eq:disph:q}
\end{equation}
where we used $q_i \equiv \langle q \rangle (\boldsymbol r)$. 
By replacing $d\boldsymbol r_j'$ in Eq. \eqref{eq:smoothed_f:derivation}
with Eq.  \eqref{eq:disph:volume_element},
we have the gradient of $\langle f\rangle$:
\begin{equation}
\langle \nabla f \rangle (\boldsymbol r_i) = \sum_j U_j \frac{f_j}{q_j}
        \nabla W_{ij}(h_i).  \label{eq:disph:smoothed_f:derivation:sum}
\end{equation}
We adopt Eqs. \eqref{eq:disph:smoothed_f:sum} and \eqref{eq:disph:q} as the basis of
DISPH.  We derive the equations of motion and energy from this basis. We note
that DISPH is also Galilei invariant.

One might think that the use of $U$ for the calculation of the volume element
would cause some inconsistency, since $U$ is not a conserved quantity. The mass
of a particle is constant, and thus looks safer.  In the following, we show that
we can construct a consistent set of equations using $U$, and that it has many
advantages over the standard SPH and that it retains important characteristics
such as force symmetry and energy conservation.  We first derive the energy
equation and then equation of motion.  We then discuss the formulation for the
estimate of the density and the implementation of the artificial viscosity.

\subsection{Energy Equation} \label{sec:disph:energy}

We need an expression of $\nabla \cdot \boldsymbol v$ to
derive the energy equation.  We start with
\begin{equation}
\nabla (q \boldsymbol v) = \nabla q \boldsymbol v + q \nabla \cdot \boldsymbol
v, \label{eq:disph:qflux}
\end{equation}
which is obtained by replacing $\rho$ in Eq. \eqref{eq:ssph:massflux} with $q$.
By applying Eq. \eqref{eq:disph:smoothed_f:derivation:sum} to Eq.
\eqref{eq:disph:qflux}, we obtain an analogy of Eq. \eqref{eq:ssph:divv}:
\begin{align}
q_i \nabla \cdot \boldsymbol v_i &= 
\sum_j U_j \boldsymbol v_j \cdot \nabla W_{ij}(h_i) 
-\boldsymbol v_i \cdot \sum_j U_j \nabla W_{ij}(h_i) \notag \\
&= -\sum_j U_j \boldsymbol v_{ij} \cdot \nabla W_{ij}(h_i).
        \label{eq:disph:divergence_v}
\end{align}
The energy equation is then given by
\begin{equation}
\frac{d u_i}{dt} = \sum_j U_j \frac{P_i}{\rho_i q_i} \boldsymbol v_{ij}
        \cdot \nabla \tilde{W}_{ij}.  \label{eq:disph:energy:u}
\end{equation}
Equation \eqref{eq:disph:energy:u} contains $\rho_i$ since $u$ is the energy per
unit mass.  The equation for $U_i$ is obtained by multiplying Eq.
\eqref{eq:disph:energy:u} by $m_i$:
\begin{equation}
\frac{d U_i}{dt} = \frac{m_i}{\rho_i} \sum_j \frac{U_j P_i}{q_i} \boldsymbol v_{ij} 
\cdot \nabla \tilde {W}_{ij}.
\end{equation}
Here, $m_i/\rho_i$ is the volume associated with particle $i$
which can be replaced by $U_i/q_i = (\gamma-1)U_i/P_i$. Thus, we have
\begin{equation}
\frac{d U_i}{dt} = (\gamma-1) \sum_j \frac{U_i U_j}{q_i} \boldsymbol v_{ij} 
\cdot \nabla \tilde {W}_{ij}.
        \label{eq:disph:energy:U}
\end{equation}

\subsection{Equation of Motion} \label{sec:disph:motion}

From the energy equation, Eq. \eqref{eq:disph:energy:U}, we derive the equation
of motion. The change in the internal energy of particles $i$ and $j$ due to
their relative motion is  
\begin{equation}
\frac{dU_i}{dt} + \frac{dU_j}{dt} = (\gamma-1) U_i U_j \left ( \frac{1}{q_i} +
\frac{1}{q_j} \right ) \boldsymbol v_{ij} \cdot 
\nabla \tilde{W}_{ij}. \label{eq:disph:energy:change:U}
\end{equation}
This change is the same as the change of the kinetic energy of particles with an
opposite sign. Thus, we have 
\begin{equation}
\frac{m_i m_j}{m_i + m_j} \boldsymbol v_{ij} \cdot
\left (\frac{d \boldsymbol v_i}{dt} - \frac{d \boldsymbol v_j}{dt} \right) 
= - \left (\frac{dU_i}{dt} + \frac{dU_j}{dt} \right). \label{eq:disph:energy:change:K}
\end{equation}
Substituting Eq. \eqref{eq:disph:energy:change:U} into Eq.
\eqref{eq:disph:energy:change:K}, we obtain
\begin{equation}
\left (\frac{d \boldsymbol v_i}{dt} - \frac{d \boldsymbol v_j}{dt} \right) 
= -(\gamma-1) \frac{m_i + m_j}{m_i m_j} U_i U_j \left ( \frac{1}{q_i} + \frac{1}{q_j} \right ) 
\nabla \tilde{W}_{ij}. \label{eq:disph:energy:change:KU}
\end{equation}
Since the motion of the center of mass of two particles is unchanged by the
interaction of two particle, we have 
\begin{equation}
\frac{d}{dt} ( m_i \boldsymbol v_i + m_j \boldsymbol v_j) = 0.
\end{equation}
Thus, we have 
\begin{equation}
m_i \frac{d \boldsymbol v_i}{dt} = 
-(\gamma-1) U_i U_j \left ( \frac{1}{q_i} + \frac{1}{q_j} \right ) 
\nabla \tilde{W}_{ij}, \label{eq:disph:euler:ij}
\end{equation}
as the contribution of particle $j$ to the equation of motion of particle $i$.

The equation of motion for particle $i$ is obtained by taking summation over
neighbor particles:
\begin{equation}
m_i \frac{d \boldsymbol v_i}{dt} = 
-(\gamma-1) \sum_j U_i U_j \left ( \frac{1}{q_i} + \frac{1}{q_j} \right )
\nabla \tilde{W}_{ij}. \label{eq:disph:euler}
\end{equation}

The right-hand side of Eq. \eqref{eq:disph:euler} contains only the energy $U$ and
energy density $q$. Thus, as far as $q$ is smooth, Eq. \eqref{eq:disph:euler} is
likely to be well-behaved.  The equation of motion of the standard SPH [Eq.
\eqref{eq:ssph:euler}] requires that both $P$ and $\rho$ are smooth.  Thus, in
our formulation, there is nothing special about the contact discontinuity. We
can therefore expect that the treatment of the contact discontinuity is
improved.  We will see this in \S \ref{sec:disph:pressure}.

Note that Eq. \eqref{eq:disph:euler} is mathematically equivalent to the
equation of motion  obtained by \citet{RitchieThomas2001}, while the derivation
is completely different.  \citet{RitchieThomas2001} started from Eq.
\eqref{eq:disph:q} and density estimate $\rho = mq/U$, but still tried to use
standard SPH estimate of Eq. \eqref{eq:smoothed_f:sum}. In order to eliminate
$\rho$ from equation of motion, they used the following formal relationship
\begin{equation}
\frac{\nabla P}{\rho} =  \frac{\nabla P}{\rho} + \frac{P}{\rho} \nabla 1,
\end{equation}
and formal identity
\begin{equation}
\nabla 1 = \sum_j m_j \frac{1}{\rho_j} \nabla W_{ij}(h) \simeq 0.
\end{equation}
Thus, their derivation was a heuristic modification of the standard SPH and they
did not employ the volume element $(\gamma-1) m_j u_j/P_j$ explicitly. We have
shown that by choosing $(\gamma-1) m_j u_j/P_j$ as the volume element, we can
derive a consistent set of SPH equations naturally.

\subsection{Artificial Viscosity} \label{sec:disph:viscosity}

To deal with shocks, the standard SPH needs an artificial viscosity term.  DISPH
also needs an artificial viscosity term.  We utilize artificial viscosity terms
which are widely used in simulations with the standard SPH.

The viscosity term for the equation of motion is 
\begin{equation}
m_i \frac{d^2 \boldsymbol r_i}{dt^2} = - m_i \sum_j m_j \Pi_{ij}
\nabla \tilde{W}_{ij}, \label{eq:visc:motion}
\end{equation}
and the corresponding form of it for the energy equation is 
\begin{equation}
\frac{d U_i}{dt} = \frac{m_i}{2} \sum_j m_j \Pi_{ij} 
\boldsymbol v_{ij} \cdot \nabla \tilde{W}_{ij}, \label{eq:visc:energy}
\end{equation}
where $\Pi_{ij}$ is the function of the strength of the artificial viscosity.

There are two types of artificial viscosity term, $\Pi_{ij}$, which are commonly
used. The most commonly used one \citep{Lattanzio+1985} is
\begin{equation}
\Pi_{ij} = 
\begin{cases}
\frac{-\alpha c_{ij} \mu_{ij} + \beta \mu_{ij}^2}{\rho_{ij}}
& \boldsymbol v_{ij} \cdot \boldsymbol r_{ij} < 0, \\
0 & \boldsymbol v_{ij} \cdot \boldsymbol r_{ij} \ge 0, \label{eq:visc:pi}
\end{cases}
\end{equation}
where $\alpha$ and $\beta$ are the control parameters for the strength of the
artificial viscosity, $c_{ij}$ is the arithmetic average of the sound speeds of
particles $i$ and $j$, $\rho_{ij} = 0.5(\rho_i + \rho_j)$, and 
\begin{equation}
\mu_{ij} = \frac{h_{ij} \boldsymbol v_{ij} 
\cdot \boldsymbol r_{ij}}{r_{ij}^2 + \epsilon h_{ij}^2}. \label{eq:visc:mu}
\end{equation}
The constant $\epsilon$ is introduced to avoid the divergence and its fiducial
value is $\sim 0.01$.

The other one is proposed by \citet{Monaghan1997} from the analogy of the
Riemann solver:
\begin{equation}
\Pi_{ij} = 
\begin{cases}
-\frac{\alpha}{2} \frac{v_{ij}^{\rm sig} w_{ij}}{\rho_{ij}}
& \boldsymbol v_{ij} \cdot \boldsymbol r_{ij} < 0, \\
0 & \boldsymbol v_{ij} \cdot \boldsymbol r_{ij} \ge 0, \label{eq:visc:pi2}
\end{cases}
\end{equation}
where $v_{ij}^{\rm sig} = c_i + c_j -3 w_{ij}$ and $w_{ij} = \boldsymbol
v_{ij}\cdot \boldsymbol r_{ij}/r_{ij}$.

Since we have the density estimate $\rho = q/u$, we have 
\begin{equation} 
\rho_{ij} = \frac{1}{2} \left( \frac{q_i}{u_i} + \frac{q_j}{u_j} \right),
\label{eq:visc:wrong_rho}
\end{equation}
for our formulation.  However, this modification of $\rho_{ij}$ leads to
unstable behavior under strong shocks. It seems to be safe to use the smoothed
mass densities of particles $i$ and $j$ evaluated using Eq.
\eqref{eq:ssph:density}.  The viscosity relates to the inertial force.
Therefore, the use of the matter distribution is reasonable. We use the smoothed
mass density to evaluate the artificial viscosity term.  In \S
\ref{sec:explosion}, we investigate the effect of the choice of the averaged
density in the artificial viscosity term. It is also  safer to use the smoothed
mass density, when one use our SPH for the simulation with a radiative cooling
term.

We use the standard Balsara switch \citep{Balsara1995} to suppress the shear
viscosity.  It is given by
\begin{equation}
F_i^{\rm Balsara} = \frac{|\nabla \cdot \boldsymbol v_i|}
{|\nabla \cdot \boldsymbol v_i| 
+ |\nabla \times \boldsymbol v_i| 
+ \epsilon_b c_i/h_i}, \label{eq:visc:balsara}
\end{equation}
and $\Pi_{ij}^{\rm Balsara} = 0.5 (F_i^{\rm Balsara}+F_j^{\rm Balsara}) \Pi_{ij}$.  Here $\epsilon_b$ is a
small value (typically $10^{-4}$).  The rotation of velocity in the standard SPH
is found in literature \citep[e.g.,][]{Monaghan1992}.  The rotation of velocity
in our SPH is calculated as follows:
\begin{equation}
\nabla \times \boldsymbol v_i =
\frac{1}{q_i} \sum_j U_j \boldsymbol v_{ij} \times \nabla W_{ij}(h_i). \label{eq:disph:rotatoni}
\end{equation}

\subsection{Grad-h Term} \label{sec:disph:grad-h}

To obtain a consistent formulation with the variable kernel size, we have to
take into account not only the gradient of kernel and physical quantities
respect to $r$ but also that respect to $h$.  Here, we take the contribution of
the variable kernel size into account.  This was accomplished by using the
Lagrangian formulation \citep{SpringelHernquist2002, Rosswog2009, Springel2010,
Hopkins2013}. Since the variation of $h$ is the first order term, the
contribution of this term is rather limited.  Indeed, we find that the
contribution of this term is prominent only in extremely strong shock problems
like the Sedov problem (See \S \ref{sec:explosion}).

We start from the Lagrangian:
\begin{equation}
L(\boldsymbol {\dot Q}, \boldsymbol Q) 
= \sum_i \frac{1}{2} m_i \boldsymbol {\dot Q}_i^2 - \sum_i U_i(\boldsymbol Q),
\label{eq:gradh:Lagrangian}
\end{equation}
where $\boldsymbol Q \equiv ({\boldsymbol r_1},{\boldsymbol r_2},,,{\boldsymbol
r_N},h_1,h_2,,,h_N)$.
We adopt the following constraint:
\begin{equation}
\phi_i = \frac{4\pi}{3} (2 h_i)^3 \frac{q_i}{U_i} - N_{\rm ngb} = 0.
\end{equation}
This constraint means that the spherical region with the radius $2 h_i$ covers
a volume of $N_{\rm ngb} \Delta V_i$.

The Euler-Lagrange equation with a constraint is as follows:
\begin{equation}
\frac{d}{dt} \frac{\partial L}{\partial \boldsymbol {\dot Q_i}}
- \frac{\partial L}{\partial \boldsymbol Q_i} = 
\sum_j \lambda_j \frac{\partial \phi_j}{\partial \boldsymbol Q_i}.
\label{eq:gradh:ELeq}
\end{equation}
First, we solve this equation regarding the kernel size.  Since the Lagrangian
and constraint do not have the first order derivative of the kernel size, Eq.
\eqref{eq:gradh:ELeq} is
\begin{equation}
-\frac{\partial L}{\partial h_i} = \sum_j \lambda_j \frac{\partial
\phi_j}{\partial h_i}.
\label{eq:gradh:ELeq:hi}
\end{equation}
The left-hand-side of this equation is 
\begin{align}
- \frac{\partial L}{\partial h_i} 
&= \frac{\partial U_i}{\partial h_i}, \notag\\
&= \frac{\partial U_i}{\partial \Delta V_i} 
\frac{\partial \Delta V_i}{\partial q_i} 
\frac{\partial q_i}{\partial h_i}, \notag\\
&= \frac{P_i U_i}{q_i^2} \frac{\partial q_i}{\partial h_i},
\label{eq:gradh:ELeq:hi:left}
\end{align}
where we used the first law of thermodynamic in an adiabatic state, $dU = -P
dV$.

The right-hand-side of Eq. \eqref{eq:gradh:ELeq:hi} is
\begin{align}
\sum_j \lambda_j \frac{\partial \phi_j}{\partial h_i} 
&= \sum_j \lambda_j 
\frac{\partial}{\partial h_i} 
\left [ \frac{4\pi}{3} (2 h_j)^3 \frac{q_j}{U_j} - N_{\rm ngb} \right ], \notag\\
&= \lambda_i 
32\pi h_i^2 \frac{q_i}{U_i} 
\left ( 1+ \frac{h_i}{3 q_i}
\frac{\partial q_i}{\partial h_i} \right ).
\label{eq:gradh:ELeq:hi:right}
\end{align}

Hence, we have 
\begin{equation}
\lambda_i = 
\frac{3 P_i}{32\pi h_i^3}\frac{U_i^2}{q_i^2} \psi_i,
\end{equation}
where
\begin{equation}
\psi_i = 
\frac{h_i}{3 q_i}
\frac{\partial q_i}{\partial h_i}
\left ( 1+ \frac{h_i}{3 q_i}
\frac{\partial q_i}{\partial h_i} \right )^{-1}.
\end{equation}

Next, let's solve the Euler-Lagrange equation regarding $\boldsymbol r$.
According to Eq. \eqref{eq:gradh:ELeq}, the equation of motion is 
\begin{align}
m_i \frac{d v_i}{dt} &=
- \nabla_i \sum_j U_j + \sum_j \lambda_j \nabla_i \phi_j \notag \\
&= - \sum_j \frac{\partial U_j}{\partial \Delta V_j} 
\nabla_i \Delta V_j \notag\\
&\quad+ \sum_j
\frac{3 P_j}{32\pi h_j^3}\frac{U_j^2}{q_j^2} \psi_j
\nabla_i \left [ \frac{4\pi}{3} (2 h_j)^3 \frac{q_j}{U_j} - N_{\rm ngb}
\right] \notag\\
&= -\sum_j \frac{P_j U_j}{q_j^2} \left (1 - \psi_j \right ) \nabla_i q_j\notag\\
&= -\sum_j \frac{P_j U_j}{q_j^2} f_j^{\rm grad} \nabla_i q_j,
\label{eq:gradh:ELeq:EoM}
\end{align}
where
\begin{equation}
f_j^{\rm grad} \equiv  \left (1 - \psi_j \right ) = 
 \left (1+ \frac{h_j}{3 q_j}
\frac{\partial q_j}{\partial h_j} \right )^{-1}.
\label{eq:gradh:ELeq:f}
\end{equation}
By using 
\begin{align}
\nabla_i q_j &= \nabla_i \sum_k U_k W_{jk} (h_j) \notag\\ 
&= U_i \nabla_i W_{ij} (h_j)
+ \delta_{ij} \sum_k U_k \nabla_i W_{ik} (h_i), \notag\\ 
\end{align}
equation \eqref{eq:gradh:ELeq:EoM} becomes 
\begin{align}
m_i \frac{d v_i}{dt} &=
-\sum_j \frac{P_j U_j}{q_j^2} f_j^{\rm grad}  \notag\\
&\quad \left ( U_i \nabla_i W_{ij} (h_j)
+ \delta_{ij} \sum_k U_k \nabla_i W_{ik} (h_i) \right ), \notag\\ 
&= -(\gamma-1) \sum_j U_i U_j \notag\\
&\quad \left (
\frac{1}{q_i} f_i^{\rm grad} \nabla_i W_{ij} (h_i)
+\frac{1}{q_j} f_j^{\rm grad} \nabla_i W_{ij} (h_j) \right).
\label{eq:gradh:ELeq:EoM:final}
\end{align}

The energy equation can be obtained as follows:
\begin{align}
\frac{d U_i}{dt} &= -P_i \frac{d \Delta V_i}{dt}, \notag\\
&= P_i \frac{U_i}{q_i^2}\frac{d q_i}{dt}.
\label{eq:gradh:ELeq:Energy}
\end{align}
Taking a time derivative of $q_i$, we have
\begin{align}
\frac{dq_i}{dt}
&= \frac{d}{dt} \sum_j U_j W_{ij}(h_i), \notag\\
&= \sum_j U_j \frac{d \boldsymbol r_{ij}}{dt} 
\cdot \nabla_i W_{ij}(h_i)
+\sum_j U_j \frac{dh_i}{dt} \frac{\partial W_{ij}(h_i)}{\partial h_i}, \notag\\
&= \sum_j U_j 
\boldsymbol v_{ij} \cdot \nabla_i W_{ij}(h_i)
+\frac{d q_i}{dt}\frac{\partial h_i}{\partial q_i}
\sum_j U_j \frac{\partial W_{ij}(h_i)}{\partial h_i}.
\end{align}
Then, 
\begin{equation}
\left ( 1 - \frac{\partial h_i}{\partial q_i}\sum_j U_j 
\frac{\partial W_{ij}(h_i)} {\partial h_i}
\right ) \frac{d q_i}{dt} 
= \sum_j U_j \boldsymbol v_{ij} \cdot \nabla_i W_{ij}(h_i).
\label{eq:gradh:ELeq:dqdt}
\end{equation}
Since $h_i^3 q_i = {\rm const}$, we have
\begin{equation}
\left ( \frac{\partial h_i}{\partial q_i} \right )^{-1} = -3 \frac{q_i}{h_i}.
\end{equation}
Therefore, Eq \eqref{eq:gradh:ELeq:dqdt} becomes 
\begin{equation}
\left ( 1 + \frac{h_i}{3 q_i} \frac{\partial q_i}{\partial h_i}
\right ) \frac{d q_i}{dt} 
= \sum_j U_j \boldsymbol v_{ij} \cdot \nabla_i W_{ij}(h_i),
\end{equation}
and using Eq. \eqref{eq:gradh:ELeq:f},
\begin{equation}
\frac{d q_i}{dt} 
= f_i^{\rm grad} \sum_j U_j \boldsymbol v_{ij} \cdot \nabla_i W_{ij}(h_i).
\label{eq:gradh:ELeq:dqdt2}
\end{equation}
By substituting Eq. \eqref{eq:gradh:ELeq:dqdt2} into Eq.
\eqref{eq:gradh:ELeq:Energy}, we finally obtain the energy equation,
\begin{equation}
\frac{dU_i}{dt} 
= (\gamma -1) \sum_j \frac{U_i U_j}{q_i} f_i^{\rm grad} 
\boldsymbol v_{ij} \cdot \nabla_i W_{ij}(h_i)
\label{eq:gradh:ELeq:Eenergy:Final}
\end{equation}
This energy equation and the equation of motion, Eq. \eqref{eq:gradh:ELeq:EoM:final},
are also density independent. 

The Lagrangian formulation guarantees the conservation of the energy and
momentum. We note that the set of equations of energy and motion, Eqs.
\eqref{eq:disph:energy:U} and \eqref{eq:disph:euler}, also conserve the energy
and momentum, since these equations are derived by assuming the conservation of
the energy and momentum.

\subsection{Pressure in Contact Discontinuities} \label{sec:disph:pressure}

The pressure around the contact discontinuity calculated with our SPH equation,
Eq. \eqref{eq:disph:q} is shown in the bottom panels of figure
\ref{fig:pressure}.  In the case of the equal-mass particle and the fixed
neighbor number (the left panel), we can see that the jump of the pressure at
the contact discontinuity in DISPH is much smaller than that in the standard
SPH.  In the case of the constant kernel size (the middle panel), the result of
DISPH is almost flat, while that of the standard SPH has a large error.  

In these two equal-mass cases, pressure still has small jumps at the contact
discontinuity. The reason is that in both cases the distribution of particles
is asymmetric. In the high-density region, the particle separation is smaller,
resulting in small integration error.  As a result, small error appears when the
kernel contains the contribution from both low- and high-density regions.  In
the case of the equal separation of particles, there is no jump in the pressure
distribution at the contact discontinuity, as shown in the rightmost panel.

\section{A Generalized form of DISPH} \label{sec:GDISPH}

In \S \ref{sec:DISPH}, we have derived equations of energy and motion which do
not depend on the density in the right-hand-side of them.  To derive these
formulation, we used $q$ instead of $\rho$ as a basis of the formulation. In an
ideal fluid, $q$ is identical to $P$ except the factor of $\gamma -1$ and the
pressure is the continuous quantity across contact discontinuities.

As we will show in the next section, this formulation works quite well in
many tests, such as shock tube test, Kelvin-Helmholtz and Rayleigh-Taylor
instability tests.  However, this formulation is not good for the extremely
strong shock tests where the pressure jump at the shock is very large.

In this section, we generalize DISPH by using an arbitrary function of pressure,
instead of $q = P/(\gamma-1)$.  If we choose the form of the function that
depends weakly on the pressure, such formulation might work well even for the
extremely strong shock. As we show in \S \ref{sec:explosion}, we found that it
indeed works well.  In the following, we derive a generalized form of DISPH.

\subsection{Equations of a Generalized DISPH}

We start from the following relation:
\begin{equation}
y_i = G(P_i), \label{eq:GDISPH:y}
\end{equation}
where $G(\boldsymbol P_i)$ is an arbitrary function of $P_i$. Formally, any kind
of function is possible as $G$.  When we introduce a physical quantity, $Z$, we
can define a new volume element as
\begin{equation}
\Delta V_i = \frac{Z_i}{y_i}. \label{eq:GDISPH:Z} 
\end{equation}

By applying Eq. \eqref{eq:GDISPH:Z} to Eq. \eqref{eq:smoothed_f},
we have
\begin{equation}
f_i = \sum_j Z_j \frac{f_j}{y_j} W_{ij}(h_i). \label{eq:GDISPH:smoothed_f:sum}
\end{equation}
When we assume $f = y$, we have
\begin{equation}
y_i = \sum_j Z_j W_{ij}(h_i). 
\label{eq:GDISPH:smoothed_y}
\end{equation}
The quantity $y$ given by Eq. \eqref{eq:GDISPH:smoothed_y} is the basic quantity
of this generalized DISPH. This is an implicit equation for $y$ and $Z$. We will
show how to solve this equation in \S \ref{sec:GDISPH:yz}. Here, we continue the
deviation of equations.  The first derivative of the physical quantity $f$ is
given by
\begin{equation}
\langle \nabla f \rangle (\boldsymbol r) = \sum  Z_j \frac{f_j}{y_j}
\nabla W_{ij}(h_i). 
\label{eq:GDISPH:smoothed_f:derivation}
\end{equation}

The equations of energy and motion are 
\begin{align}
\frac{d U_i}{dt} &= \sum_j \frac{P_i Z_i Z_j}{y_i^2}  \boldsymbol v_{ij} 
\cdot \nabla \tilde {W}_{ij}, \label{eq:GDISPH:energy} \\
m_i \frac{d \boldsymbol v_i}{dt} &= 
-\sum_j Z_i Z_j \left ( \frac{P_i}{y_i^2} + \frac{P_j}{y_j^2} \right )
\nabla \tilde{W}_{ij}, \label{eq:GDISPH:euler}
\end{align}
where we use the following relation:
\begin{equation}
\nabla (y \boldsymbol v) = \nabla y \boldsymbol v + y \nabla \cdot \boldsymbol
v, \label{eq:GDISPH:yflux}
\end{equation}
which can be rewritten as
\begin{align}
\nabla \cdot \boldsymbol v_i 
&= -\frac{1}{y_i} \sum_j Z_j \boldsymbol v_{ij} \cdot \nabla W_{ij}(h_i).
        \label{eq:GDISPH:divergence_v}
\end{align}

The rotation of velocity in this generalized DISPH is expressed as
\begin{equation}
\nabla \times \boldsymbol v_i =
\frac{1}{y_i} \sum_j Z_j \boldsymbol v_{ij} 
\times \nabla W_{ij}(h_i). \label{eq:GDISPH:rotatoni}
\end{equation}

The time derivative of $Z$ is obtained from Eq. \eqref{eq:GDISPH:Z}.
By taking the time derivative of Eq. \eqref{eq:GDISPH:Z}, we have 
\begin{align}
\frac{d Z_i}{dt} &= Z_i \left ( 
\frac{1}{\Delta V_i} \frac{\partial \Delta V_i}{\partial t} 
+ \frac{1}{y_i} \frac{\partial y_i}{\partial t} \right ), \notag\\
&= Z_i \left ( 
\frac{1}{\Delta V_i} \frac{\partial \Delta V_i}{\partial t} 
+ \frac{1}{y_i} \frac{\partial \Delta V_i}{\partial t} \frac{\partial y_i}{\partial \Delta V_i} \right ), \notag\\
&= \frac{Z_i}{\Delta V_i} \frac{\partial \Delta V_i}{\partial t} 
\left ( 1+ \frac{\Delta V_i}{y_i} 
\frac{\partial y_i}{\partial \Delta V_i} \right ).
\end{align}
Adopting the following relations: 
\begin{align}
\frac{1}{\Delta V_i} \frac{\partial \Delta V_i}{\partial t}
&= \nabla \cdot \boldsymbol v_i, \\
\frac{d \log P_i}{d \log \Delta V_i} &= -\gamma, \\
\frac{d \log y_i}{d \log P_i} &= \frac{d \log G(P_i)}{d \log P_i} \equiv \zeta(P_i), 
\end{align}
we finally obtain 
\begin{align}
\frac{d Z_i}{dt} &= 
Z_i (\zeta(P_i) \gamma-1) \frac{1}{\Delta V_i} 
\frac{\partial \Delta V_i}{\partial t}, \notag\\
&= (\zeta(P_i) \gamma-1) \sum \frac{Z_i Z_j}{y_i} \boldsymbol v_{ij} \cdot \nabla
W_{ij}(h_i).
\label{eq:GDISPH:dZdt}
\end{align} 

Note that these equations is reduced to those of DISPH shown in \S
\ref{sec:DISPH}, if we choose $q_i$ as $y_i$. Thus, the equations of DISPH
obtained in \S \ref{sec:DISPH} is a special case of the generalized DISPH. In
this special case, the time derivative of $Z$ is identical to the energy
equation.

\subsection{A Generalized DISPH with the Grad-h Term}

We can obtain the grad-h term by using the new volume element Eq.
\eqref{eq:GDISPH:Z}.  We begin the derivation by the Lagrangian, Eq.
\eqref{eq:gradh:Lagrangian}, and a constraint with the new volume element of the
generalized DISPH,
\begin{equation}
\phi_i = \frac{4\pi}{3} (2 h_i)^3 \frac{y_i}{Z_i} - N_{\rm ngb} = 0.
\label{eq:GDISPH:Lagrangian:Constraint}
\end{equation}

By solving the Euler-Lagrange equation, we finally obtain the following set 
of equations:
\begin{align}
m_i \frac{d \boldsymbol v_i}{dt} &= 
-\sum_j Z_i Z_j \notag\\
&\quad \left (
\frac{P_i}{y_i^2} f_i^{\rm grad} \nabla_i W_{ij} (h_i)
+\frac{P_j}{y_j^2} f_j^{\rm grad} \nabla_i W_{ij} (h_j) \right), 
\label{eq:GDISPH:Lagrangian:Momentum} \\
\frac{dU_i}{dt} &= 
\frac{P_i Z_i}{y_i^2} f_i^{\rm grad} \sum_j Z_j \boldsymbol v_{ij} \cdot \nabla_i W_{ij}(h_i),
\label{eq:GDISPH:Lagrangian:Energy}
\end{align}
where 
\begin{equation}
f_i^{\rm grad} = \left (1+ \frac{h_i}{3 y_i}
\frac{\partial y_i}{\partial h_i} \right )^{-1}.
\label{eq:GDISPH:Lagrangian:f}
\end{equation}
The time derivative of $Z$ is as follows:
\begin{equation}
\frac{d Z_i}{dt} = 
(\zeta(P_i) \gamma-1) f_i^{\rm grad} 
\sum \frac{Z_i Z_j}{y_i} \boldsymbol v_{ij} \cdot \nabla W_{ij}(h_i).
\end{equation} 
Again, these equations are reduced to those obtained in \S
\ref{sec:disph:grad-h}, if we adopt $q_i$ as $y_i$.

\subsection{Solving the Implicit Relation Between $y$ and $Z$}
\label{sec:GDISPH:yz}

We have the energy equation [Eq. \eqref{eq:GDISPH:energy}] and the equation
\eqref{eq:GDISPH:dZdt} for $Z$. If we integrate both $U$ and $Z$, they will
become inconsistent due to the truncation error of the integration scheme. Thus,
we should ``correct'' the value of $Z$ so that it is consistent to $U$.  To
solve this matter, we use an iteration method to obtain the consistent value of
$Z$ with the equation of state. The procedure is as follows:
\begin{enumerate}
\item Predict $Z$ at the next step, $Z_{i}^{\rm new,p}$, by calculating
\begin{equation}
Z_{i}^{\rm new,p} = Z_{i}^{\rm old} + \frac{d Z_i}{dt} dt,
\label{eq:GDISPH:Z_new:p}
\end{equation}
where $Z_{i}^{\rm old}$ is the value of $Z$ at the last step.  \item Using this
$Z_{i}^{\rm new,p}$, we obtain $y_{i}^{\rm new}$ by evaluating Eq.
\eqref{eq:GDISPH:y}.
\item Then, we evaluate the value of $Z$ using $y$:
\begin{equation}
\hat Z_{i}^{\rm new} = (y_{i}^{\rm new})^{1-\frac{1}{\zeta(P_i)}} (\gamma-1) U_i.
\label{eq:GDISPH:Z_new:hat}
\end{equation}
\item We update the value of $Z$:
\begin{equation}
Z_{i}^{\rm new} = Z_{i}^{\rm old} + \zeta(P_i) (\hat Z_{i}^{\rm new}-Z_{i}^{\rm old}).
\label{eq:GDISPH:Z_new}
\end{equation}
\item We again calculate $y$ using Eq. \eqref{eq:GDISPH:y}:
\begin{equation}
y_{i}^{\rm new} = \sum Z_{i}^{\rm new} W_{ij}(h_i).
\label{eq:GDISPH:y_new}
\end{equation}
\end{enumerate}
We repeat this iteration from the procedure 3 if necessary.

Since
\begin{equation}
Z \propto y^{1-\frac{1}{\zeta(P_i)}},
\end {equation}
the error in $y$ is amplified by the direct substitution when $\zeta(P_i) <
1/2$. To avoid this instability, we adopt Eq. \eqref{eq:GDISPH:Z_new}, in which
the new $Z_i$ is taken between old $Z_i$ and predicted $Z_i$, to update $Z$ in
the iteration.

\subsection{Specific Form of $G(P)$}

Formally, any kind of function is possible as $G(P)$.  However, since our aim of
the introduction of such a function for the formulation is to reduce errors
induced by huge pressure jumps, it is natural to select a function which depends
weakly on the value of pressure. Here, we consider the following power law form:
\begin{equation}
y_i = P_i^{\zeta},
\end{equation}
where $\zeta$ is a constant value less than unity.  This selection makes
$\zeta(P)$ a constant value $\zeta$.  For instance, when we choose $\zeta =
0.1$, the pressure contrast with the ten orders of magnitude is reduced that
with an orders of magnitude.  It is worth noting that $\zeta = 1$ is the special
case in which the equations are reduced to those we obtained in \S
\ref{sec:DISPH} and the Ritchie \& Thomas formulation.  We expect that this
formulation would improve behaviors under strong shocks.

It is also noted that $\zeta = 0$ is the other special case in which the
equations solve the evolution of volume elements which is independent of
pressure gradient.  The fundamental equation of this SPH is 
\begin{equation}
1 = \sum_j V_j W_{ij} (h_i).
\end{equation}
The equations of energy and motion can be drawn in the same way we have shown in
this paper.  This formulation keeps important properties of the generalized
DISPH that it is density independent formulation and it reduces the problem due
to pressure jump around strong shocks.  Further investigation of this
formulation is beyond the scope of this paper. We will study this formulation
elsewhere.

\section{Numerical Experiments} \label{sec:results}

In this section, we show the results of several standard tests for fluid
schemes, for both the standard SPH and our new SPH.  In \S \ref{sec:method}, we
describe our numerical code briefly. In \S \ref{sec:shocktube}, we show the
results of the shock tube tests. Then we show the evolution of 
system which is initially in hydrostatic equilibrium in \S \ref{sec:surface}. In
\S \ref{sec:KH} and \S \ref{sec:RT}, tests for two important fluid instabilities
are carried out.  Point like explosion tests are shown in \S
\ref{sec:explosion}.  In \S \ref{sec:blob}, we show the results of the blob
tests which was first proposed by \citet{Agertz+2007}.  As an extra test, we
describe the mixing of a two phase fluid with a solid body like spoon which is
found in \cite{Springel2010}.  In all tests, our new SPH shows much better
result compared to that of the standard SPH.

\subsection{Numerical Method} \label{sec:method}

We used {\tt ASURA}, a parallel $N$-body/SPH code, as the framework of current
numerical experiments. {\tt ASURA} adopts the leap-frog method for the
time-integration. For simplicity, we used the shared steps with variable
time-steps. The time-step is given by 
\begin{equation}
dt = \min_i dt_i, \label{eq:dt}
\end{equation}
where 
\begin{equation}
dt_i  = C_{\rm CFL} \frac{2 h_i}{\max_j v_i^{\rm sig}}, \label{eq:dti}
\end{equation}
and $C_{\rm CFL} = 0.3$.

For the standard SPH, we first evaluated the densities and kernel sizes of
particles using Eq. \ref{eq:ssph:density:ind_h} and iteration.  Then, we
calculated the pressure gradient and the time-derivative of the internal energy
using Eqs. \ref{eq:ssph:euler:ind_h} and \ref{eq:ssph:energy:ind_h}.  In DISPH,
we computed $q$ using Eq.  \ref{eq:disph:q} and kernel sizes first, and then we
calculated the pressure gradient and the time derivative of the internal
energy using Eqs. \ref{eq:disph:euler} and \ref{eq:disph:energy:U}.  For most of
tests, we used a special DISPH which adopt $q$ as a fundamental value.  We
investigated the advantages of the generalized DISPH and the grad-h term in
the strong shock test in \S \ref{sec:explosion}.
We used Eq.  \ref{eq:visc:pi2} as the artificial viscosity term in both cases
and we adopted $\alpha = 1$ as a fiducial value.  In the loop where we compute
$q$, we also compute smoothed mass densities since they are used in the
calculation of the artificial viscosity term.  The Balsara switch was also
applied.  To avoid the pairing instability, we used a first derivative of the
kernel which has a cuspy core \citep{ThomasCouchman1992}.  Note that this
modification leads to an inconsistent sound speed and other quantities within $s
= |\boldsymbol r - \boldsymbol r'|/h < 2/3$ \citep[See footnote 8
in][]{Price2012Review}.  The essential solution is to use kernels which do not
show the pairing instability, for instance, the kernel proposed by
\citet{Read+2010} and Wendland kernels \citep{DehnenAly2012}.

The kernel size of each particle is determine to keep the number of neighbor
particles within the range of $32 \pm 2$. As an exception, in the
one-dimensional tests shown in \S \ref{sec:shocktube}, the kernel size is
evaluated by 
\begin{equation}
h = \eta \left ( \frac{m}{\rho} \right ),
\end{equation}
where $\eta = 1.2$ for the Sod's shock tube tests and $\eta = 2.4$ for the
strong shock tube tests. For DISPH, we used the smoothed density for the
evaluation of the kernel size.

\subsection{Shock Tube Tests} \label{sec:shocktube}

The Sod shock tube \citep{Sod1978} is the most basic test for numerical
schemes for compressible fluid. This test shows the shock capturing ability of
schemes.  In SPH, not only the profile of the shock front but also the behavior
of the contact discontinuity is important. Here, we show the results of one- and
three-dimensional shock tube tests.

The setup is as follows.  We prepared the periodic domain of $-1 \le x < 1$ for
the one-dimensional tests and $-1 \le x < 1$, $-1/16 \le y < 1/16$, and $-1/16
\le z < 1/16$ for the three dimensional tests.  The initial condition is given
by 
\begin{equation}
\begin{cases}
\rho = 1, P = 1, v = 0, &  x < 0, \\
\rho = 0.25, P = 0.1795, v = 0, & x\ge0.
\end{cases}
\end{equation}
To express this initial condition, we use equal-mass particles and place 800 and
200 particles in the left and right domains, respectively, regularly in the
one-dimensional tests.  In three dimensional tests, we place 40000 and 10000
particles in the left and right domains, respectively, and a glass-like
particle distribution was used.  We set $\gamma = 1.4$ and gave the internal
energy to each particle to ensure the given $P$.

In addition to the Sod shock tube, we performed a one-dimensional strong shock
test.  The initial condition for this test is given by 
\begin{equation}
\begin{cases}
\rho = 1, P = 1000, v = 0, &  x < 0, \\
\rho = 1, P = 0.01, v = 0, & x\ge0.
\end{cases}
\end{equation}
We use 1000 equal-mass particles in the computational domain of $-1 \le x < 1$
with the equal separation.

\begin{figure}
\begin{center}
\epsscale{1.0}
\plotone{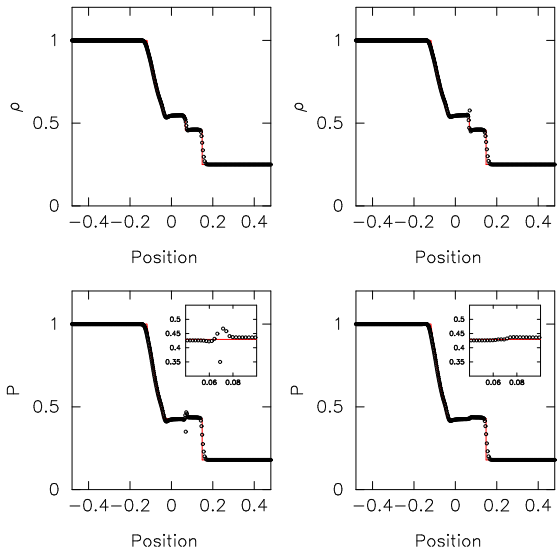}
\caption{The results of the one dimensional shock tube tests for the standard
SPH and DISPH at $t = 0.1$.  Density (upper row) and pressure (bottom row)
are shown.  Circles indicate the physical quantities of each SPH particle, while
red curves represent the analytic solutions. Insets in the pressure panels are
the close-up views around the contact discontinuity.  
\label{fig:1dst}
}
\end{center}
\end{figure}

Figure \ref{fig:1dst} shows the results of the one-dimensional shock tube
tests with the standard SPH and DISPH. The density (upper row) and pressure
(bottom row) of each particle are plotted by circles. The red curves represent
the analytic solutions.

The standard SPH reproduces the analytic solution of the density distribution
well. The shock front is resolved by $\sim 7$ particles. The jump of the
density at the contact discontinuity is resolved by a similar number of
particles.  The pressure shows large variations near the contact discontinuity,
though it should be constant.  Since Eq. \ref{eq:ssph:euler} of the
standard SPH contains a large error near the contact discontinuity, in order to
achieve zero acceleration, pressures of particles must have large variations.
This result is the same as the results of previous works with the standard SPH
\citep[e.g.,][]{Springel2005, Price2008}.

In DISPH, unlike the case of the standard SPH, the pressure around the
contact discontinuity does not show a large jump. The reason is simply that the
energy density is used instead of the mass density. The energy density is
constant at the contact discontinuity.  The reason why there is a small change
in the pressure is that the particle separation changes at the contact
discontinuity. As we showed in figure \ref{fig:pressure}, our new SPH still has
small error in the pressure, due to the finite number of particles in the
kernel. This error caused the change in the pressure in figure \ref{fig:1dst}.

\begin{figure}
\begin{center}
\epsscale{1.0}
\plotone{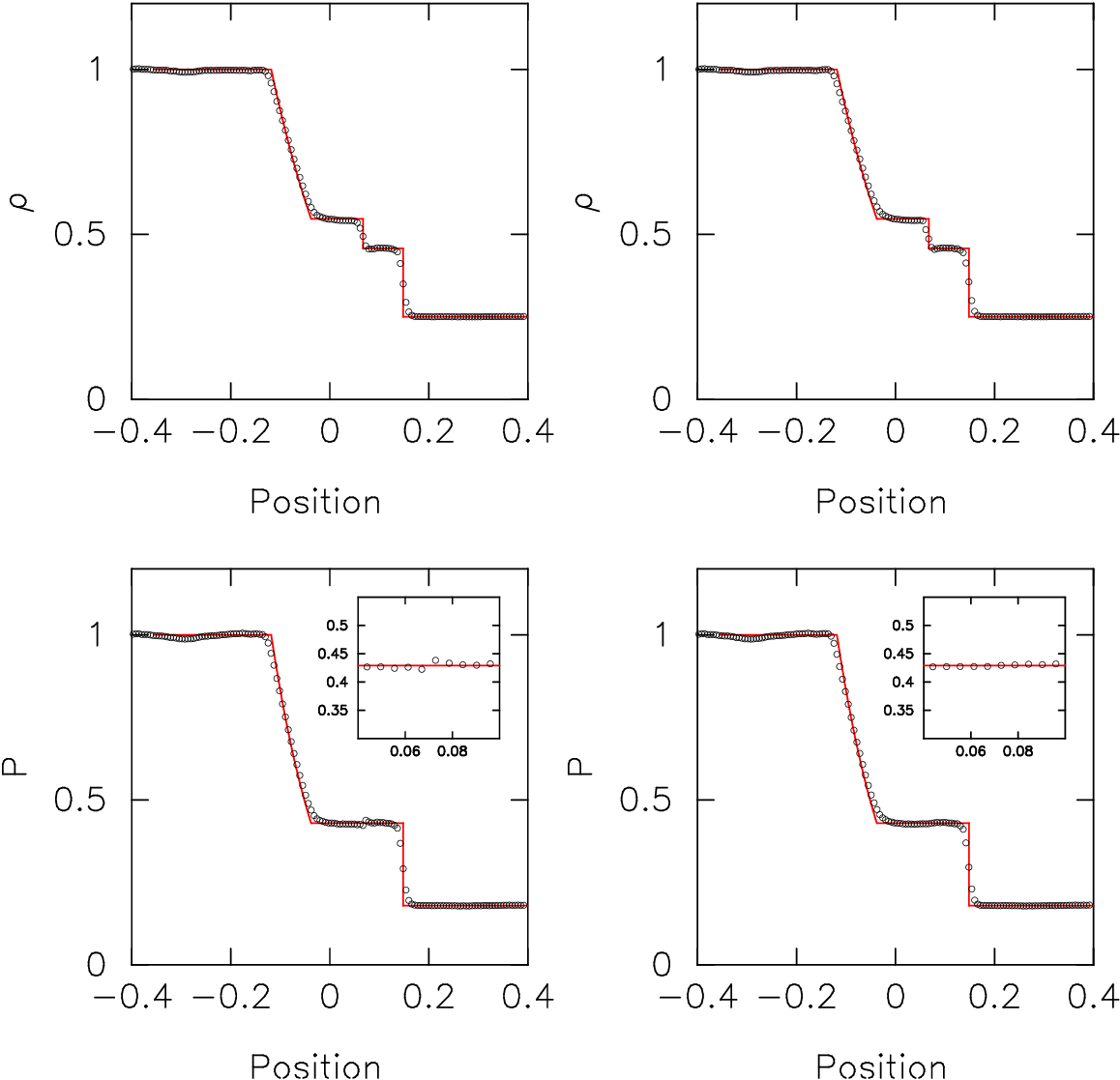}
\caption{The results of the three dimensional shock tube tests for the standard
SPH and DISPH at $t = 0.1$.  Density (upper row) and pressure (bottom row)
are shown.  Circles indicate the averaged physical quantities of SPH particles,
while red curves represent the analytic solutions.
\label{fig:3dst}
}
\end{center}
\end{figure}

The results of the three dimensional shock tube tests for the standard SPH and
our SPH are shown in figure \ref{fig:3dst}.  In this figure, the circles
represent average values of particles in bins with the width of the mean
particle separation at the high density part.  Again, we can see a variation in
the pressure around the contact discontinuity in the case of the standard SPH.
In the case of our SPH, there is no such variation.

\begin{figure}
\begin{center}
\epsscale{1.0}
\plotone{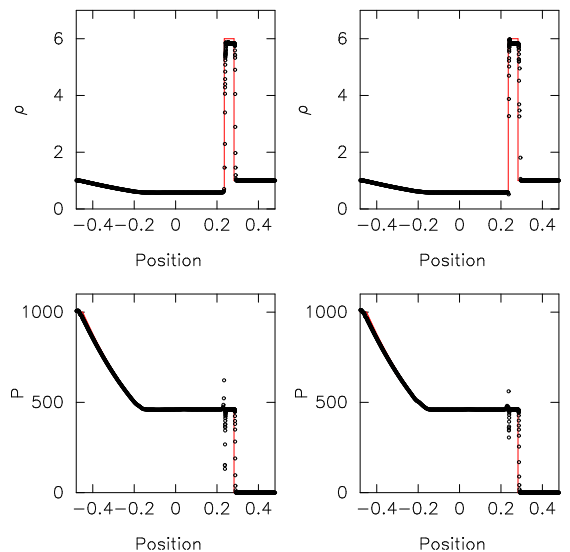}
\caption{The same as figure \ref{fig:1dst}, but for the strong shock tube tests
at $t = 0.012$.
\label{fig:1dst:strong}
}
\end{center}
\end{figure}

Figure \ref{fig:1dst:strong} shows the results of the strong shock tube tests
for the standard SPH and DISPH. The shock front and the contact discontinuity in
the density distribution is well reproduced in the both cases.  In this extreme
test, both runs show jumps in the pressure distribution around the contact
discontinuity.  The absolute value of the pressure jump in DISPH is much
smaller than that in the standard SPH.  The jump found in the pressure in our
SPH is caused by the asymmetry in the particle distribution (see \S
\ref{sec:disph:pressure}).  Overall, DISPH can handle such a strong shock
problem, even when a very large pressure jump exists initially.  This result is
quite reassuring.  In DISPH, it is assumed that pressure is smooth, which is not
a valid assumption at the shock front. Thus, it could fail to capture very
strong shocks.  The result shown in figure \ref{fig:1dst:strong} shows that is
not the case and new SPH can handle very strong shocks.

\subsection{Hydrostatic Equilibrium Tests} \label{sec:surface}

As is shown in \S \ref{sec:SSPH}, in the standard SPH, particles feel {\it
unphysical repulsive force} at the interface of the contact discontinuity.
Therefore, in order to establish the hydrostatic equilibrium, the distance
between particles at the different sides of the contact discontinuity must
become larger than the ``true'' value. What is the consequence of this repulsive
force?  Here, we show the result of a simple test which helps us to understand
the problem of the unphysical repulsive force.  Similar test has been used in
\citet{HessSpringel2010}.

We follow the evolution of two fluids with different values of density, but with
the same pressure.  We performed two-dimensional tests. The computational
domain is a square of the unit size, $0 \le x < 1$ and $0 \le y < 1$, with a
periodic boundary condition.  Initial conditions are 
\begin{align}
\rho &=
\begin{cases}
4 &  0.25 \le x \le 0.75~{\rm and}~0.25 \le y \le 0.75, \\
1 & {\rm otherwise}, 
\end{cases} \\
P&=2.5, \\
\gamma &= 5/3.
\end{align}
We tried two different realizations. In the first one, the particle mass is the
same for the entire computational region. Thus, the inter-particle distance is
smaller in the high density region. In the second one, particles in the high
density region is four times more massive than particles in the low-density
region. In both cases, particles are initially in a regular grid.  For the
equal-mass case, the number of particles in the dense region is 4096 and that in
the ambient is 3072.  For the equal-separation cases, those are 1024 and 3072,
respectively.  Initial velocities of particles were set to zero.  Since the
system is initially in the hydrostatic equilibrium, particles should not move,
except for small local adjustments.

\begin{figure*}[htb]
\begin{center}
\epsscale{1.0}
\plotone{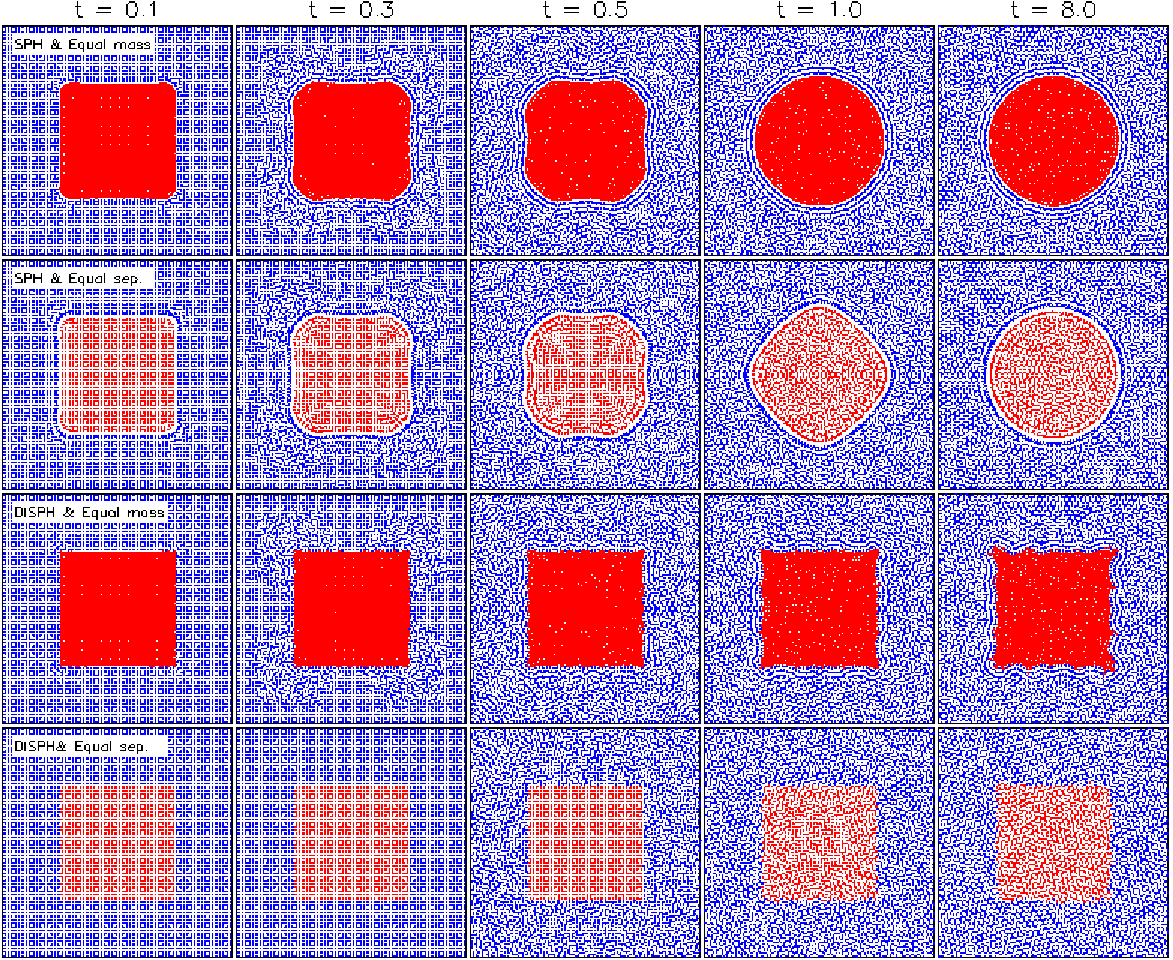}
\caption{Snapshots of a two-fluid system at $t = 0.1$, $0.3$, $0.5$, $1$ and
$8$.  The red and blue points indicate the positions of particles with $\rho =
4$ and $\rho = 1$, respectively.  The upper two rows are the results of the
standard SPH, while the lower two rows are those of DISPH.  The first and
third rows show the results of the equal-mass cases, whereas the second and
fourth rows show those of the equal separation and unequal mass cases.
\label{fig:surface}
}
\end{center}
\end{figure*}

Figure \ref{fig:surface} shows the time evolution up to $t = 8$.  There is a
clear difference between the result of the standard SPH and that of DISPH.
With the standard SPH, the high-density region, which initially has a square
shape, quickly becomes rounder and almost completely circular by $t = 8$.  We can
understand this unphysical rounding as follows.  As we stated in \S
\ref{sec:SSPH} and \S \ref{sec:disph:pressure}, unphysical repulsive force
between particles operates at the contact discontinuity. We can see the effect
of this force in the development of the gap of the distribution of particles
near the boundary of two fluids.  Because of this gap, the bulk of the system is
slightly compressed.  The system seeks to achieve the energy minimum, by
minimizing the surface area of the contact discontinuity.  Thus, the
high-density region evolves to a circular shape, which minimizes the length of
the boundary.  In other words, the repulsive force effectively adds the
``surface tension''.

DISPH gives a far better solution, as we can see in the lower two rows of figure
\ref{fig:surface}. The overall square shape remains there till  the end of the
simulation in the equal-mass case.  The result of the unequal-mass case is even
better.  The equation of motion of DISPH eliminates the unphysical surface
tension completely.

\begin{figure}[htb]
\begin{center}
\epsscale{1.0}
\plotone{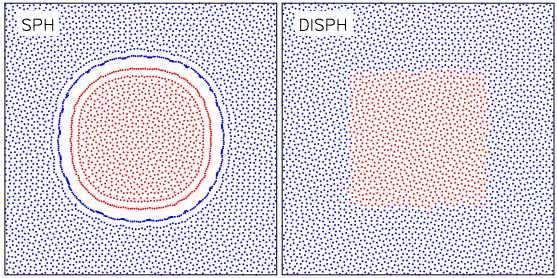}
\caption{The final state ($t = 8$) of a two fluid system with the density
contrast of 64.  The red and blue points are the positions of particles with
$\rho = 64$ and $1$, respectively. The particle separation is constant and the
particle mass difference is 1:64.
\label{fig:surface:1:64}
}
\end{center}
\end{figure}

Figure \ref{fig:surface:1:64} shows the final state of the two-fluid system
with the density contrast of 64.  Our SPH handles the system without any problem
(right panel).  On the other hand, in the calculation with the standard SPH, a
wide and empty ring structure is formed between two fluids.

\subsection{Kelvin-Helmholtz Instability Tests} \label{sec:KH}

After the work by \citet{Agertz+2007} which demonstrated clearly that the
standard SPH cannot deal with the Kelvin-Helmholtz instability correctly, many
researchers proposed modifications of SPH to solve the problem (see \S
\ref{sec:intro}).  In this section, we investigate how DISPH handles the
Kelvin-Helmholtz instability.

We prepared a two-dimensional computational domain,
$0 \le x < 1$ and $0 \le y < 1$.
The periodic boundary condition was used.
The initial density is 
\begin{equation}
\rho =
\begin{cases}
1 (\equiv \rho_l) & 0 \le y < 0.25,~0.75 \le y < 1, \\
2 (\equiv \rho_h) & 0.25 \le y < 0.75.
\end{cases}
\end{equation}
We used equal-mass particles.  The numbers of particles in the high and
low dense regions are 131072 and 65522, respectively.  We set $P = 2.5$ and
$\gamma = 5/3$.  The high and low density regions had the initial velocities of
$v_{x,h} = 0.5$ and $v_{x,l} = -0.5$ in the $x$ direction, respectively.

We have used $N_{\rm nb} = 32\pm2$ as the neighbor number. This value might seem
a bit large, but we found it guarantees the good sampling of the particles in
the low-density region at the interface.  When we used $N_{\rm nb} = 16 \pm 2$,
the variation of the pressure at the interface becomes too large. For the
artificial viscosity, we used $\alpha =1$ with the Balsara switch.

We added a small velocity perturbation to the particles near the interfaces,
following \citet{Price2008}. The velocity perturbation in the $y$ direction is
as follows:
\begin{equation}
\Delta v_y = 
\begin{cases}
A \sin [-2 \pi (x+0.5)/\lambda], & |y-0.25| < 0.025 \\
A \sin [2 \pi (x+0.5)/\lambda], & |y-0.75| < 0.025,
\end{cases}
\end{equation}
where $\lambda = 1/6$ and $A=0.025$.

The time-scale of the growth of the Kelvin-Helmholtz instability is 
\begin{equation}
\tau_{\rm kh} = \frac{\lambda (\rho_h + \rho _l)}{\sqrt{\rho_h \rho_l}
|v_{x,h}-v_{x,l}|}.
\end{equation}
For our test setup, $\tau_{\rm kh} = 0.35$. We followed the evolution up to $t =
8 \tau_{\rm kh}$.

\begin{figure*}[htb]
\begin{center}
\epsscale{1.0}
\plotone{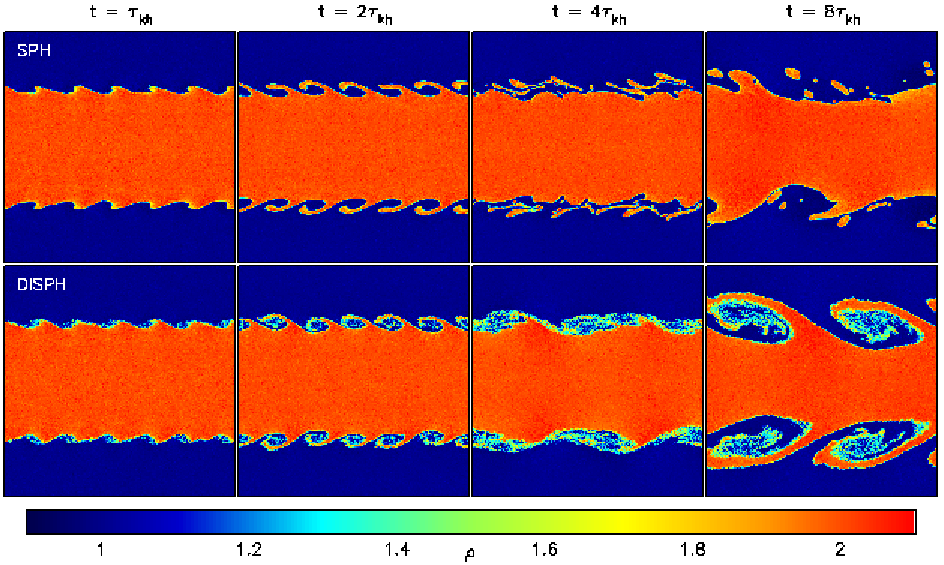}
\caption{The density maps from the two dimensional shear flow test at $t = 1, 2,
4$ and $8~\tau_{\rm kh}$. The upper panels show the results of the standard SPH,
while the bottom panels show those of DISPH. The color code of the density is
given at the bottom.
\label{fig:2DKH:Snap}
}
\end{center}
\end{figure*}

The results are shown in figure \ref{fig:2DKH:Snap}.  The difference between two
results is clear. In the run with the standard SPH, perturbations grow till $t =
\tau_{\rm kh}$, but the unphysical surface tension inhibited the growth of
roll-like structures. The stretched high-density fluids break apart ($t = 4
\tau_{\rm kh}$) and form blobs ($t = 8 \tau_{\rm kh}$). This evolution is
completely different from those obtained by Euler codes
\citep[e.g.,][]{Agertz+2007}.  On the other hand, DISPH shows a very good result
which is comparable to those with Euler codes and with SPH codes adopting the
\citet{RitchieThomas2001} equation of motion or the artificial conductivity
\citep[see ][]{Price2008}.  \citet{Price2008} reported that the instability grew
but did not develop prominently when the Ritchie \& Thomas formulation was used.
He argued that this failure was because of noise when the Ritchie \& Thomas
formulation was used and the low resolution (See figure 5 in his paper).  In our
test, we observed extended Kelvin-Helmholtz instabilities with the resolution
same as that used in \citet{Price2008}.  When we turned off the Balsara switch,
instabilities did not developed, and we obtained the result similar to Price's
result. The use of the Balsara switch is important for such a simulation of
shear flows.

\begin{figure*}[htb]
\begin{center}
\epsscale{1.0}
\plotone{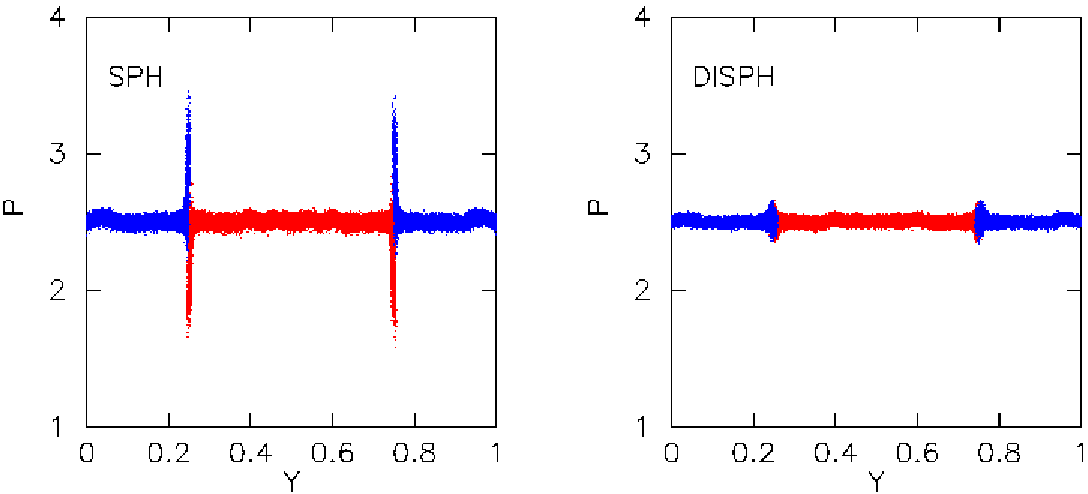}
\caption{Pressure of each particle along the $y$ direction at $t =
0.4~\tau_{kh}$.  The left panel shows the result of the standard SPH, whereas the right
panel shows that of DISPH. Particles initially in the high- (low-) density
region are expressed with red (blue) points.
\label{fig:2DKH:Pressure}
}
\end{center}
\end{figure*}

Figure \ref{fig:2DKH:Pressure} shows the cross section of the pressure
distribution along the $y$-axis. We can see that a very large pressure jump
exists around the contact interfaces, in the case of the standard SPH.
The surface tension at the interface of the two fluids prevents the normal
growth of the Kelvin-Helmholtz instability.  On the other hand, there is no such
jump in the case of DISPH.  Since the pressure and particle distribution is
well-behaved at the interface, the growth of the Kelvin-Helmholtz instability is
not suppressed.

\subsection{Rayleigh-Taylor Instability Tests} \label{sec:RT}

\citet{Abel2011} demonstrated that the standard SPH cannot follow the
development of the Rayleigh-Taylor instability correctly.  We show the result
with our SPH as well as that with the standard SPH.

\begin{figure}[htb]
\begin{center}
\epsscale{1.0}
\plotone{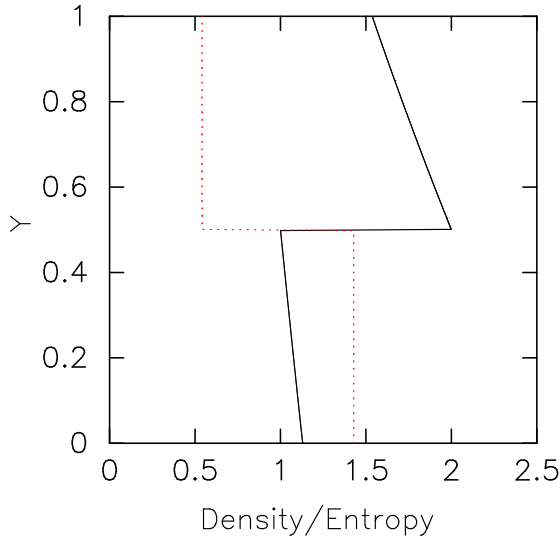}
\caption{Initial distributions of density and entropy in the vertical direction.
Solid and dotted curves indicate density and entropy, respectively.
\label{fig:2DRT:Init}
}
\end{center}
\end{figure}

The initial setup is as follows.  We prepared the two dimensional computational
domain of $0 \le x < 1$ and $0 \le y < 1$. We placed two fluids separated at $y
= 0.5$. The density just above (below) the interface was set to $\rho_h \equiv 2$
($\rho_l \equiv 1$). These two fluids were initially in the hydrostatic
equilibrium.  Further, we assumed that each fluid was initially isoentropic. The
density distributions of these fluids in the vertical direction are given by
\begin{equation}
\rho =
\begin{cases}
\rho_l \left [ 1+\frac{\gamma-1}{\gamma} \frac{\rho_l g (y-0.5)}{P_0} \right
]^{\frac{1}{\gamma-1}} & y < 0.5, \\
\rho_h \left [ 1+\frac{\gamma-1}{\gamma} \frac{\rho_h g (y-0.5)}{P_0} \right
]^{\frac{1}{\gamma-1}} & y \ge 0.5,  \label{eq:rt:init}
\end{cases}
\end{equation}
where $g=-0.5$ is the gravitational constant, $P_0 = 10/7$ is the value of
pressure at the interface, and $\gamma = 1.4$. The initial density and entropy
profiles are shown in figure \ref{fig:2DRT:Init}.  To ensure the initial density
distribution given by Eq. \eqref{eq:rt:init}, we first placed equal-mass particles
on the regular grid with the separation of 1/512. Then, we adjusted the vertical
separation of each particle set having the same $y$ to reproduce the density
distribution. The particle mass was set to $5.7\times10^{-6}$ and the total
number of particles was 247296.  The periodic boundary condition was imposed on
the $x$ direction.  Particles with $y < 0.1$ and $y>0.9$ were fixed at the
initial positions and they were not allow to change their internal energy.

The velocity perturbation in the vertical direction was added as the seed of the
instabilities.  We carried out runs with two kinds of the seed.
For the first test, we added the velocity perturbation to
particles in the range of $0.3 < y < 0.7$, and the form of the perturbation is 
\begin{equation}
\Delta v_y(x,y) = \delta_{vy} [1 + \cos(4 \pi x)]\{1 + \cos[5 \pi (y- 0.5)]\}.
\end{equation}
We set $\delta_{vy} = 0.025$.  For the second test, we added the velocity
perturbation of the form:
\begin{equation}
\Delta v_y(x,y) = \sum_{j=20}^{40} a_j \frac{n_j}{k_j} \cos (k_j x) \exp(-0.05 k_j |y-0.5|),
\end{equation}
and 
\begin{equation}
n_j^2 = k_j |g| \left ( \frac{\rho_h-\rho_l}{\rho_h+\rho_l} \right ),
\end{equation}
where $n_j$ is the linear growth rate of the Rayleigh-Taylor instability, and
$k_j = j \pi/L(\equiv1)$ is the wave number of the perturbation.  The amplitude
of each mode, $a_j$, was drawn from a Gaussian distribution with the variance of
unity at random. This initial velocity perturbation is based on
\citet{Youngs1984} with slight modifications. Velocities of the particles
outside the perturbed region was set to zero. We call these two tests
single-mode and multi-mode tests, respectively.

\begin{figure*}[htb]
\begin{center}
\epsscale{1.0}
\plotone{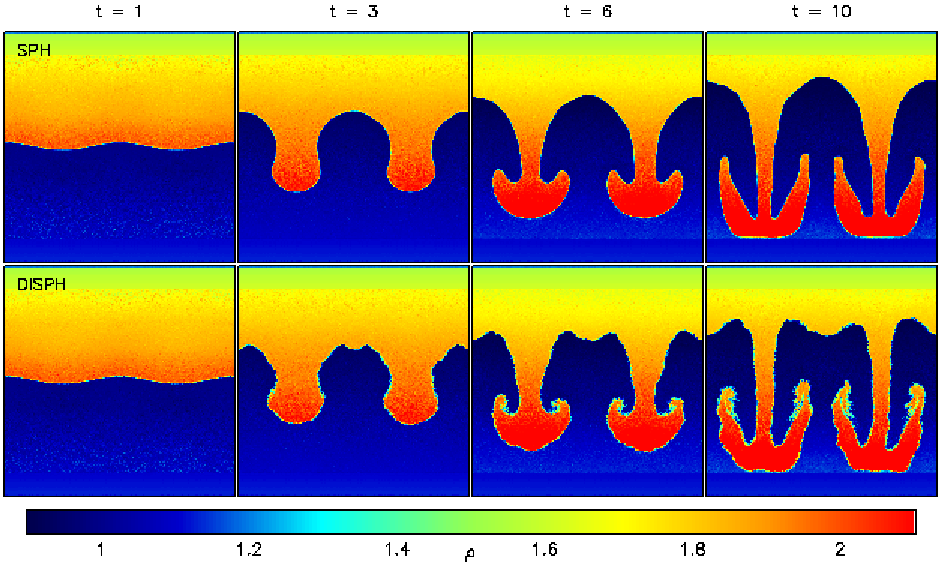}
\caption{The density maps of the two dimensional Rayleigh-Taylor instability
tests at $t = 0.5, 3, 4$ and $5$. The upper panels show the results of the
standard SPH, while the bottom panels show those of DISPH. The color code of the
density is given at the bottom.
\label{fig:2DRT:Snap}
}
\end{center}
\end{figure*}

In figure \ref{fig:2DRT:Snap}, the growth of the Rayleigh-Taylor instability in
the case of the single-mode test is shown. The Rayleigh-Taylor instability develops
in calculations with both of the standard SPH and our SPH, but the structures of
them are quite different.  The unphysical surface tension of the standard SPH
again prevents the development of the fine structures on the surface of the
Rayleigh-Taylor fingers. Thus, the result looks quite different from those
obtained with Euler schemes.  On the other hand, in the calculation with our
SPH, the overall evolution of the Rayleigh-Taylor instability in our SPH shows
excellent agreement with those with Euler schemes and the moving mesh scheme
\citep[see][]{Springel2010}.

\begin{figure*}[htb]
\begin{center}
\epsscale{1.0}
\plotone{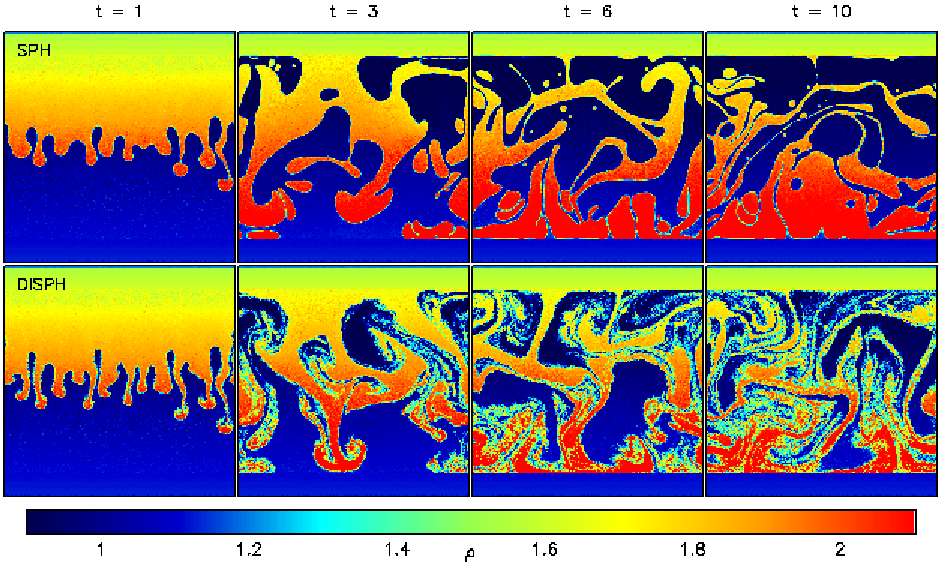}
\caption{The same as figure \ref{fig:2DRT:Snap:Random}, but for the multi-mode
perturbations.
\label{fig:2DRT:Snap:Random}
}
\end{center}
\end{figure*}

Figure \ref{fig:2DRT:Snap:Random} shows the growth of the Rayleigh-Taylor
instability with the multi-mode perturbations with the standard SPH and DISPH.
The global phase mixing of fluids can be seen in the result with our SPH.  On
the other hand, due to the unphysical surface tension, the mixing is
significantly suppressed in that of the standard SPH. The distribution of two
fluids looks like a mixture of oil and water.

\subsection{Point Like Explosion Tests} \label{sec:explosion}

In this section we describe the results of the test calculations for a
point-like explosion.

We prepared a three-dimensional computational domain of $0\le x,y,z<1$ with
a periodic boundary condition. Then, we placed $64^3$ equal-mass particles in
that domain and made a glass like distribution. The particle mass is $1/64^3$.
Thus the initial density was unity.  The explosion energy was injected to the
center of the domain.  The total thermal energy of unity was distributed
following the shape of the SPH kernel with the kernel size in which $N_{\rm s}$
particles were included. Here, we tested $N_{\rm s} = 32$ and $N_{\rm s} = 256$.
After that, we added the thermal energy of $10^{-6}$ of the particle with the
maximum energy to all particles. This energy difference between the hot region
and the ambient corresponds to the supernova explosion in a cold cloud.  In
this test, we adopted the viscosity parameter $\alpha = 3$.

We compared results of three schemes, {\it i.e.,} the standard SPH and our SPH
with $q$ and $y=P^{\zeta}$.  For DISPH with $y = P^{\zeta}$, we adopted $\zeta
= 0.1$ and the three iterations to determine $Z$ in each time-step.  We
investigated the effects of the grad-h term, the adopted value of $N_{\rm s}$,
and the use of the density derived from EOS for the artificial viscosity term.
The equations of motion and energy with the grad-h term for the standard SPH are
Eqs.  (25) and (22) in \citet{Springel2010Review}, while those for our SPH using
$q$ are Eqs.  \eqref{eq:gradh:ELeq:EoM:final} and
\eqref{eq:gradh:ELeq:Eenergy:Final}.  For the generalized DISPH, we used Eqs.
\eqref{eq:GDISPH:euler} and \eqref{eq:GDISPH:energy} for the run without the
grad-h term and Eqs. \eqref{eq:GDISPH:Lagrangian:Momentum} and
\eqref{eq:GDISPH:Lagrangian:Energy} for the run with the grad-h term,
respectively.  Note that, unlike experiments in \citet{Hopkins2013}, the equations
without the grad-h term we used here is an energy conserving ones, although
these equations neglect the gradient of the kernel size. They are
different from the case with $f_i^{\rm grad} = 1$.

\begin{figure*}[htb]
\begin{center}
\epsscale{1.0}
\plotone{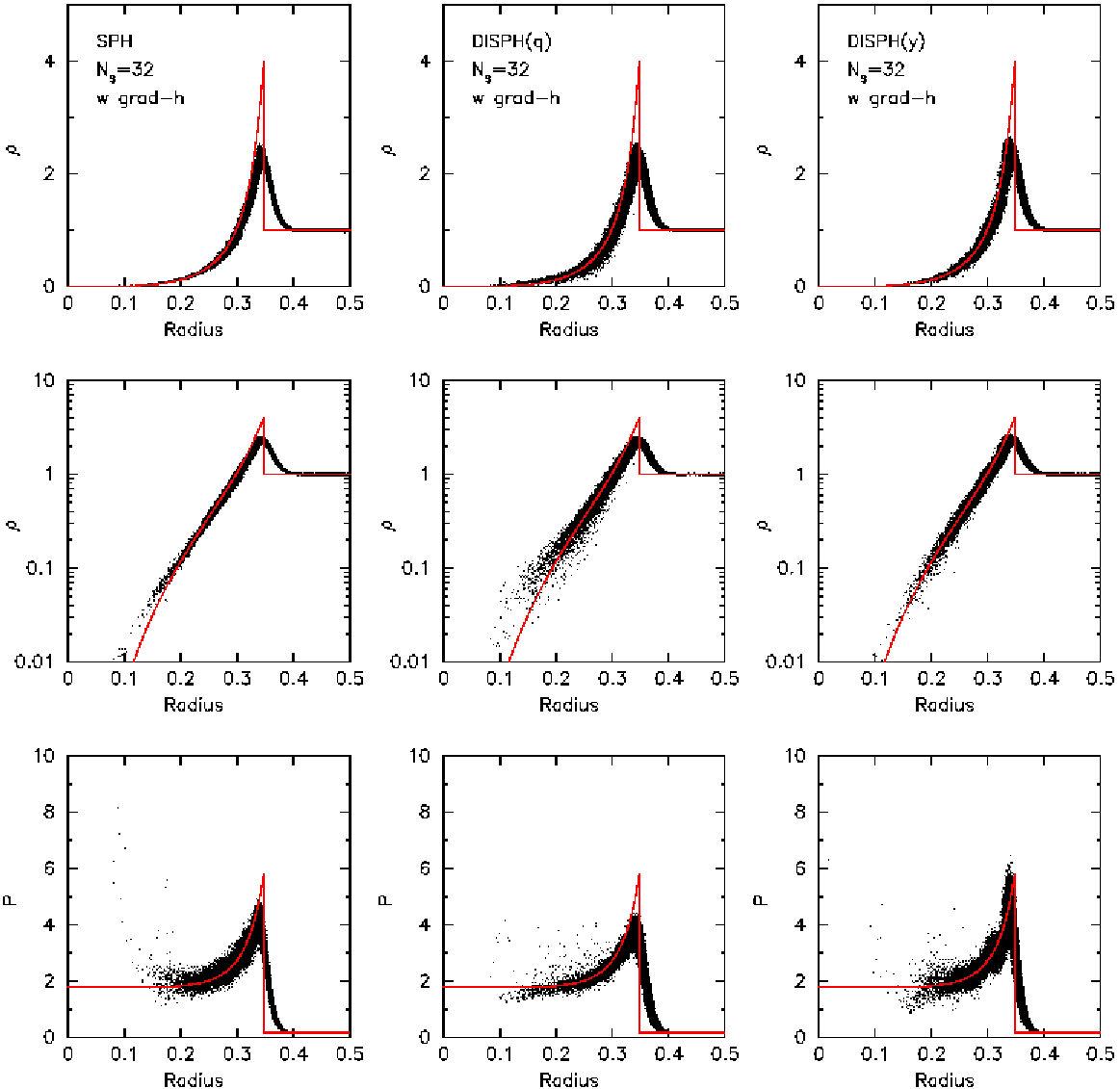}
\caption{Density and pressure profiles of three runs using the standard SPH and
DISPH with $q$ and $y$ at $t = 0.05$.  The grad-h term is taken into account
and $N_{\rm s} = 32$.  Dots represent the quantity of each particle. All
of particles are used for plots. The smoothed density is used to depict the
density profiles. Red curves are the semi-analytic solution
\citep{Sedov1959Book}.
\label{fig:Sedov}
}
\end{center}
\end{figure*}

Figure \ref{fig:Sedov} shows the density and pressure profiles of three runs.
In all runs, the grad-h term is used.  From the top row, we can see that the
density profiles of all runs are basically consistent to the analytic solution.
When we compare these results, we find that the result of the standard SPH
exhibits the smallest scatter in density. Our SPH with $y$ shows the highest
peak at the edge of the shell.

\begin{figure*}[htb]
\begin{center}
\epsscale{0.67}
\plotone{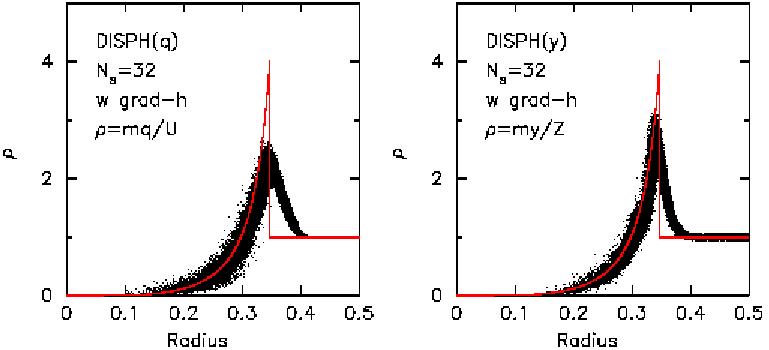}
\caption{Density profiles of two runs using DISPH with $q$ and $y$ at $t
= 0.05$. Here, the densities evaluated from the EOSs, $\rho = mq/U$ and $\rho =
my/Z$, are used.  The grad-h term is taken into account and $N_{\rm s} = 32$.
\label{fig:Sedov:RhoEOS}
}
\end{center}
\end{figure*}

These density profiles are drawn using the smoothed density even though our
schemes can obtain their intrinsic densities by using EOS.  To compare to them, 
we prepared the density profiles of our schemes which used the density
derived from EOS, {\it i.e.}, $\rho = m q/U$ and $\rho = my/Z$.  These profiles
are shown in figure \ref{fig:Sedov:RhoEOS}.  These profiles follow the analytic
solution well.  The peak of our SPH with $y$ reaches to $3$ and this value is
higher than that obtained using the smoothed density. 

However, the use of EOS density has several disadvantages. First, the scatter of
density is larger than that obtained using the smoothed density, although the
increase of the scatter is not so prominent.  Second, at very early phase, the
EOS density profile shows a large error. Indeed, the error in this test case,
the error is several orders of magnitude at the initial step.  If the radiative
cooling is included and the cooling time is comparable to or shorter than the
dynamical time, this may lead to wrong results.  If the dynamical time is the
shortest one of a system, this effect is negligible.  Thus, this depends
strongly on the situation. It would be safer to use the smoothed density for the
evaluation of the amount of radiative cooling, because the cooling rate depends
on the square of density and would be sensitive to the errors in density.

The logarithmic plots of density at the middle row in figure \ref{fig:Sedov}
indicate differences among runs more clearly.  The result of the standard SPH
shows that there is a systematic offset from the analytic solution in the
central region, since it is hard to reproduce such a very less dense region
using a given particle mass.  The worst result is provided by DISPH with
$q$. It has the largest scatter in density, and in addition, it shows the
systematic offset in the central region.  In the result of DISPH with $y$, we
found that the density profile follows the analytic solution to much less dense
region and the degree of the density scatter is less than that found in DISPH
with $q$.

\begin{figure*}[htb]
\begin{center}
\epsscale{1.0}
\plotone{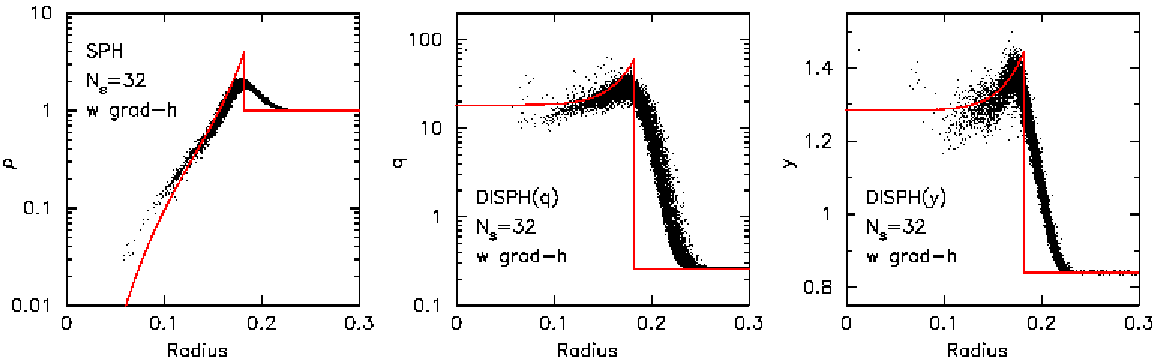}
\caption{Comparison of fundamental quantities of three runs. From left to right,
the results of the standard SPH and DISPH with $q$ and $y$ at $t = 0.01$
are shown.  Note that the vertical axis in the right panel is linear.
\label{fig:Sedov:Fundamental}
}
\end{center}
\end{figure*}

The pressure profiles are shown in the bottom row of figure \ref{fig:Sedov}.  We
can see that the standard SPH overestimates the pressure in the central, less
dense region, whereas our SPH show improved pressure profiles. This behavior
comes from the fact that our SPH adopts the pressure (energy density) or its
power as the smoothed value of the formulation.  Note however that the most
central particle in DISPH with $q$ has very large error which can observed in
figure \ref{fig:Sedov:Fundamental}.  The pressure profile of DISPH with $y$
follows the postshock profile of the analytic solution very well. Only this run
can reproduce the peak of the pressure profile.  This is because we solved the
smoothed $y = P^{0.1}$ which has quite shallow gradient in the postshock region.

Figure \ref{fig:Sedov:Fundamental}, which shows the profile of the basic
quantities {\it i.e.}, $\rho$, $q$, and $y$ at $t = 0.01$, tells us the
advantage of our SPH with $y=P^{0.1}$ over others clearly.  The density
distribution of the standard SPH in the postshock region has a change with
several orders of magnitude due to the expansion. In our SPH with $q$, a 
decrease of $q$ in the postshock region is much moderate.  
The range of $y$ in the postshock region is less than $10\%$. Since the profile
of the basic quantity is quite smooth, DISPH with $y$ can follow the peak
of the pressure profile very well.

\begin{figure*}[htb]
\begin{center}
\epsscale{1.0}
\plotone{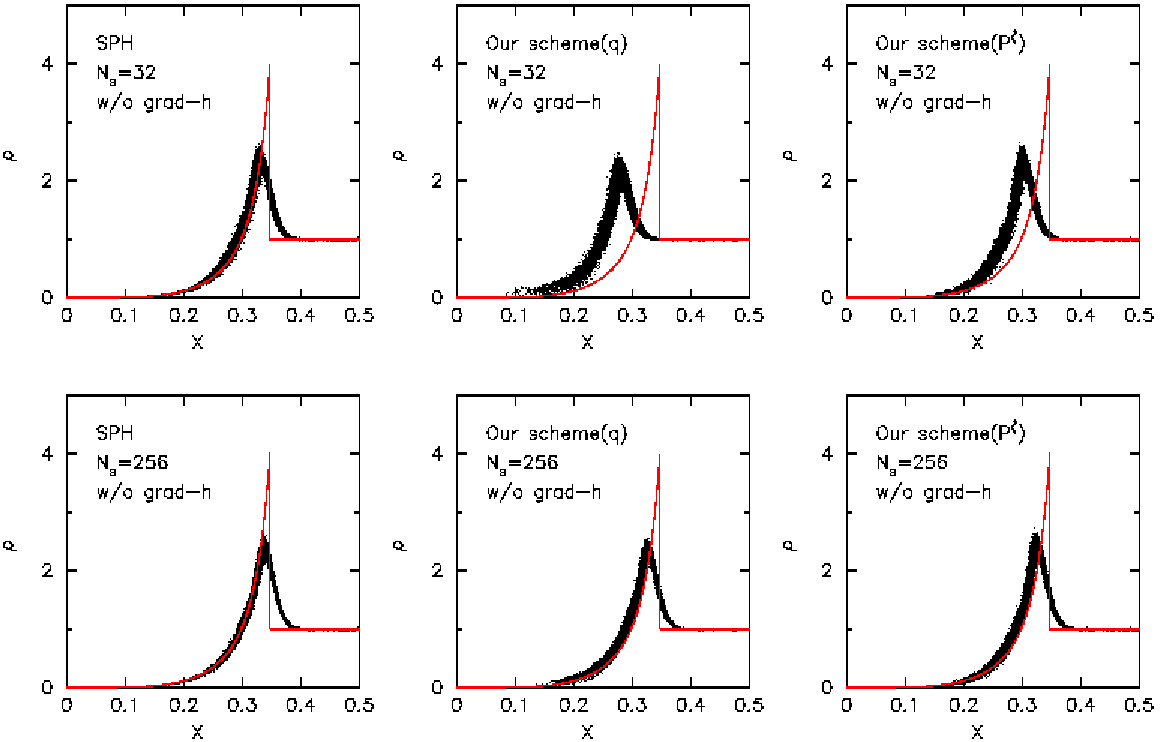}
\caption{Effects of the grad-h term and the adopted value of $N_{\rm s}$.
The top row shows density profiles without the grad-h term and $N_{\rm s} = 32$, while 
bottom row exhibits those without the grad-h term and $N_{\rm s} = 256$.
The smoothed ``mass ''density is used to draw these density profiles.  These
profiles are obtained from snapshots at $t = 0.05$.
\label{fig:Sedov:GradhNs}
}
\end{center}
\end{figure*}

The density profiles without the grad-h term are shown in figure
\ref{fig:Sedov:GradhNs}.  Even though the grad-h term is excluded, the standard
SPH can reproduce the density profile regardless of the value of $N_{\rm s}$.
This is because the value of $\partial \rho/\partial h$ in $f^{\rm grad}$ is
rather small in the standard SPH.  When $N_{\rm s} = 256$ is adopted, 
the profile becomes much smooth compared to that with $N_{\rm s} = 32$ due to the
relatively smooth, initial thermal-energy profile.  On the other hand, the
density profiles of our schemes with $N_{\rm s} = 32$ have clear delay of the
shock front. This delay can be recovered when we adopted $N_{\rm s} = 256$. In
our SPH with $q$, the gap between Eq. \eqref{eq:gradh:ELeq:f} and unity become
quite large at the early phase. Even with our SPH with $y$, the gap in Eq.
\eqref{eq:GDISPH:Lagrangian:f} and unity is non-negligible at the early phase.
These gaps lead to delay of the shock front in this test.  The amount of
the delay in our SPH with $y$ is smaller compared to that in our SPH
with $q$. This means $\partial y/\partial h < \partial q/\partial h$ and is
reasonable since we used $\zeta = 0.1$.

We show the energy errors in calculations in table \ref{tab:Sedov:Energy}.
Here, we defined the energy error as $|E(0.05)-E(0)|/E(0)$, where $E(t)$ is the
summation of the kinetic and thermal energy of all particles at the time $t$.
The energy errors in the standard SPH runs are around $10^{-4}$. In the runs
with $N_{\rm s} = 32$, the energy error in the run with the grad-h term is an
order of magnitude smaller than that without the grad-h term. However, there is
no difference in the results of the runs with $N_{\rm s} = 256$. The
contribution of the grad-h term is small for the standard SPH runs.  

The energy errors in the DISPH runs with $q$ are comparable with those in the
standard SPH runs.  However, the grad-h term plays an important role for the
DISPH runs with $q$.  When the grad-h term is used, the energy error decreases
even for $N_{\rm s} = 256$.  For the runs using DISPH with $y$, the energy
errors are an order of magnitude larger than others, however the values are
still acceptable (less than $\sim 0.1$ \%). When we increase the iteration count
from 3 to 10, the energy errors decrease by a factor of few or an order of
magnitude. The results are insensitive to the use grad-h term and the adopted
value of $N_{\rm s}$ in this case.

We note that the energy errors in the runs without the grad-h term is
sufficiently low.  This is natural since our formulation without the grad-h term
is constructed to conserve energy and momentum.  \citet{Hopkins2013} pointed
out that the runs with $f^{\rm grad} = 1$ did not conserve the total energy
\citep[see figure 2 in][]{Hopkins2013}.  However, he just adopted $f^{\rm grad}
= 1$ and did not use the symmetrized kernel in his tests.  Obviously, this
reformulation breaks conservations of energy and momentum.

\begin{table}[htb]
\caption{Energy error in the Sedov tests. 
}\label{tab:Sedov:Energy}
\begin{center}
\begin{tabular}{lrrrr}
\hline
\hline
Run & Iteration & $N_{\rm s}$ & with grad-h & without grad-h\\
\hline
SPH & N/A& $32$  & $4.8\times10^{-5}$ & $5.3\times10^{-4}$ \\
SPH & N/A& $256$  &  $3.6\times10^{-4}$ & $3.4\times10^{-4}$ \\
\hline
DISPH(q) & N/A& $32$  & $6.5\times10^{-4}$& $1.5\times10^{-3}$ \\
DISPH(q) & N/A& $256$  &  $3.9\times10^{-5}$ & $6.2\times10^{-4}$ \\
\hline
DISPH(y) &3  & $32$  & $2.7\times10^{-3}$ & $6.7\times10^{-3}$ \\
DISPH(y) &3  & $256$  &  $4.1\times10^{-3}$ & $3.5\times10^{-3}$ \\
\hline
DISPH(y) &10 & $32$  & $2.8\times10^{-4}$ & $1.1\times10^{-3}$ \\
DISPH(y) &10 & $256$  &  $1.1\times10^{-3}$ & $6.7\times10^{-4}$ \\
\hline
\end{tabular}
\end{center}
\end{table}

\begin{figure*}[htb]
\begin{center}
\epsscale{0.67}
\plotone{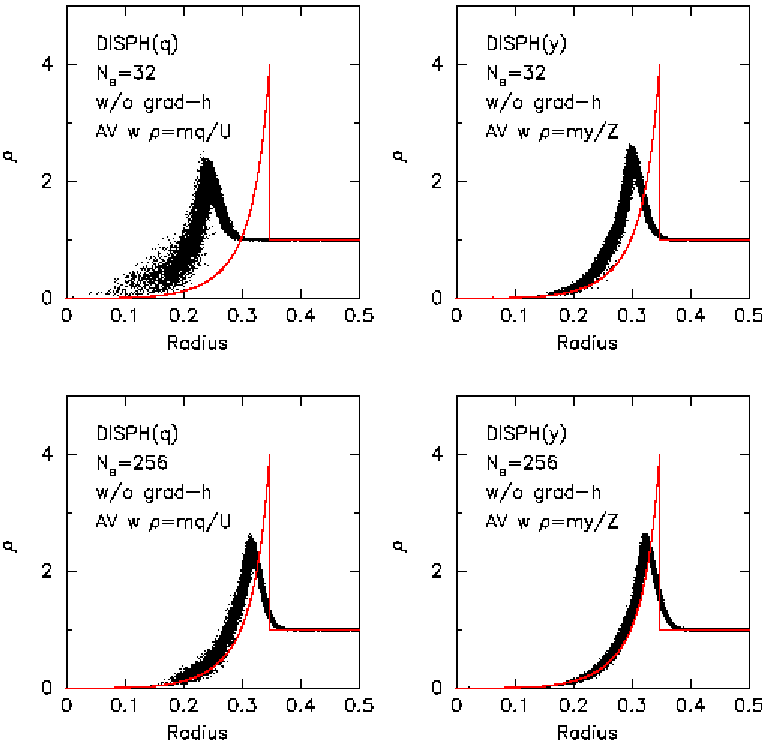}
\caption{Density profiles for DISPH with $q$ and $y$ at $t = 0.05$.
$\rho = mq/U$ and $\rho = my/Z$ is used to evaluate $\rho_{ij}$ in the
artificial viscosity term in the run of DISPH with $q$ and $y$,
respectively.
\label{fig:Sedov:WrongVisc}
}
\end{center}
\end{figure*}

Finally, we investigate the case that the density obtained by EOS is used for
the evaluation of the artificial viscosity term.  Instead of the smoothed
mass density, we adopted $\rho = mq/U$ and $\rho = my/Z$ to evaluate $\rho_{ij}$ in
Eq. \eqref{eq:visc:pi2} for DISPH with $q$ and $y$, respectively, and
depicted the density profiles in figure \ref{fig:Sedov:WrongVisc}.
By comparing the left panels in figure \ref{fig:Sedov:WrongVisc} and the middle
panels in figure \ref{fig:Sedov:GradhNs}, we can see that the use of the EOS
density makes the situation much worse. The scatter of the density profile and
the delay of the shock front increase.  We found that the delay due to the
exclusion of the grad-h term is recovered when we used $N_{\rm s} =256$.
However, in this case, there is still a non-negligible delay.  The error in the
EOS density induced by the error in pressure profile distorted the evolution of
the expanding shell.  The viscosity is the quantity which relates to the
inertial force. Thus, the use of the smoothed mass density in the artificial
viscosity term is generally suitable and this result support this idea.
Interestingly, in DISPH with $y$, the use of the EOS density for the artificial
viscosity makes no difference.

In appendix A of \cite{ReadHayfield2012}, they discussed the problem in the
Ritchie \& Thomas formulation.  They showed that it could not handle the strong
shock (see figure A1 in their figure).  We found that the treatment of the
artificial viscosity term and the exclusion of the grad-h term are keys of this
problem.  Careful choice of equations, the density evaluation method for the
artificial viscosity term, and the energy input scale provides good results even
when the SPH formulation with the smoothed pressure (energy density) is adopted.

\subsection{Blob Tests} \label{sec:blob}

In this subsection, we performed the blob test proposed by \citet{Agertz+2007}.
This test incorporates both the Kelvin-Helmholtz and Rayleigh-Taylor
instabilities.

We used Read's initial condition of the blob test \citep{Read+2010,
ReadHayfield2012} {\footnote {We obtained the initial condition from the
following URL: {\url http://www-theorie.physik.uzh.ch/astrosim/code/}}}.
The computational domain was $0 \le x < 2000~{\rm kpc}$, $0 \le y < 2000~{\rm
kpc}$, and $0 \le z < 6000~{\rm kpc}$, and the periodic boundary condition was
imposed. A cold cloud of the density $\rho_c = 3.13\times 10^{-7}$ in the mass
unit of $2.3\times10^5~M_{\sun}$and the length unit of $1~{\rm kpc}$ and
temperature $T_{\rm c} = 10^6~{\rm K}$ was centered at $(x,y,z) = (1000~{\rm
kpc},1000~{\rm kpc},2000~{\rm kpc})$. The radius of this cloud was $197~{\rm
kpc}$. This cloud was embedded in the diffuse ambient gas of which density and
temperature were $\rho_{\rm a} = 3.13 \times 10^{-8}$ and $T_{\rm a} = 10^7~{\rm
K}$, respectively.  The ambient gas had the velocity of $v_z = 1000~{\rm
km~s^{-1}}$. Thus, the Mach number of the flow to the cloud was $2.7$.  The
total number of particle for the system is 4643283.  We integrated the system up
to $t = 5 \tau_{\rm kh}$, where $\tau_{\rm kh} = 2~{\rm Gyr}$ is the typical
growth time-scale of the Kelvin-Helmholtz instability in this test
\citep{Agertz+2007}.

\begin{figure*}[htb]
\begin{center}
\epsscale{1.0}
\plotone{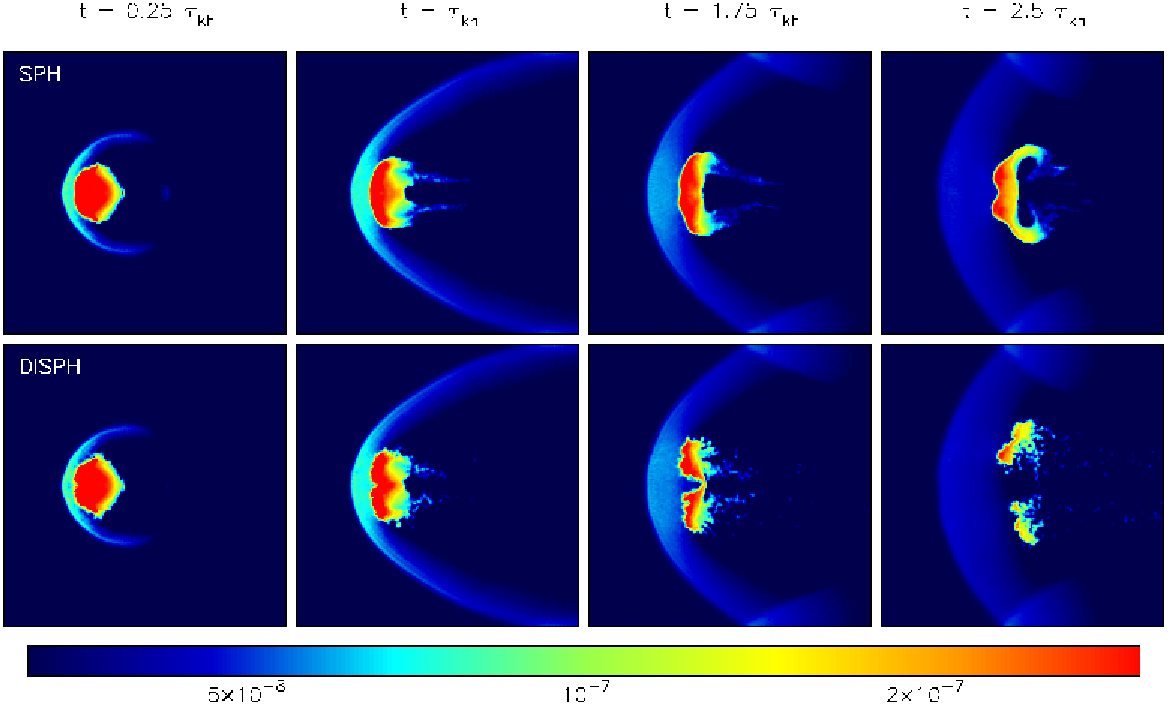}
\caption{The density maps at $t = 0.25$, $1.0$, $1.75$ and $2.5~\tau_{\rm kh}$.
The upper and lower panels show the results with the standard SPH and DISPH, respectively.
The color code of the density is given at the bottom.
\label{fig:blob:Snap}
}
\end{center}
\end{figure*}

Figure \ref{fig:blob:Snap} shows the snapshots of the cloud core.  The upper and
lower panels are the results with the standard SPH and DISPH, respectively.
Their evolutions were quite different.  The blob simulated with the standard SPH
retained the single cloud structure until the late stage of the simulation. This
behavior is consistent with those with the standard SPH shown in
\citet{Agertz+2007}.  In contrast, the blob surface was disrupted in the run
with our SPH, due to the growth of the instabilities on the surface. The blob
fragmented into several peaces and mixed eventually with the ambient gas. This
behavior is similar to those obtained by Euler codes.

\begin{figure}[htb]
\begin{center}
\epsscale{1.0}
\plotone{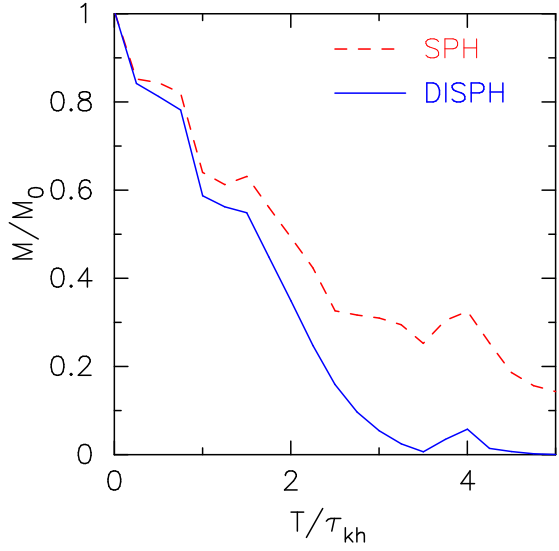}
\caption{The evolution of the blob mass up to $t = 5~\tau_{\rm kh}$.
\label{fig:blob:core}
}
\end{center}
\end{figure}

The evolution of the blob mass is shown in figure \ref{fig:blob:core}. Here we
show the mass of gas with $\rho > 0.64~\rho_{\rm c}$ and $T < 0.9~T_{\rm a}$,
following \citet{Agertz+2007}. At $t = 2.5~\tau_{\rm kh}$, the blob mass in the
run of DISPH became $\sim10~\%$ of the initial mass. This result is consistent
with the results of the Euler codes \citep[see figure 6 in][]{Agertz+2007}.  The
evolution of the blob mass in the standard SPH was much slower compared to that
in DISPH.

\subsection{Mixing of Two Phase Fluid with Spoon} \label{sec:mixing}

In this subsection, we discuss the results of simulations of two fluid mixed
with a solid body like a spoon. With this test, we can see the ability of a
particle-based scheme to handle fixed and moving boundary conditions. The setup
of this test is similar to that in section 8.9 of \cite{Springel2010}.

The initial setup is as follows. We prepared the two-dimensional computational
domain of $-0.025 \le x \le 1.025$ and $-0.025 \le y \le 1.025$. The lower part
of the domain $y < 0.6$ was the dense region with the density of $1$ while the
upper part of the domain $y > 0.6$ was the less-dense region with the density of
$0.5$.  To make this density distribution, we employed the equal-mass particles
of which mass is $1.87\times10^{-6}$ and we placed them on the regular grid with
the separation of $1.05/768$.  Then, we doubled the vertical separation of
particles in $y > 0.6$. The pressure was unity in the whole region, with $\gamma
= 5/3$.  Sound speeds in the dense and less-dense regions were $1.3$ and $1.8$
in the simulation unit, respectively.  Particles which were out of the range
$0\le x < 1$ and $0\le y < 1$ were fixed at the initial positions and initial
physical quantities. These fixed particles express the fixed boundary condition.
The velocity of all particles were initially set to zero.  The number of
particles in the dense and less-dense regions in $0\le x < 1$ and $0\le y < 1$
were $319185$ and $107604$, respectively, and that of the boundary particles
were $41064$.

\begin{figure}[htb]
\begin{center}
\epsscale{1.0}
\plotone{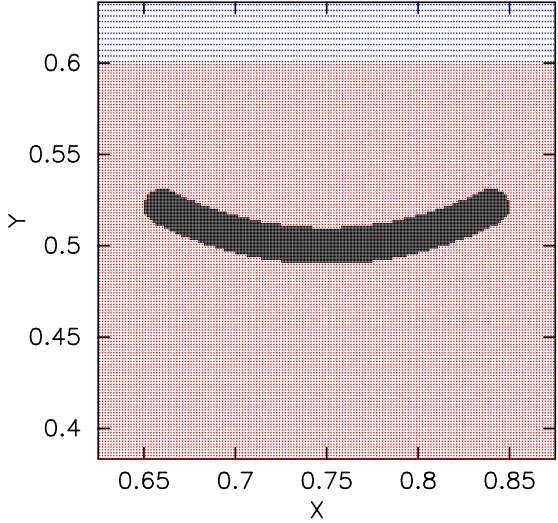}
\caption{The two-dimensional structure of the spoon expressed by SPH particles.
Black particles comprise the spoon. Red and blue particles are the fluid
particles for high and low densities, respectively.  
\label{fig:Mixing:Spoon}
}
\end{center}
\end{figure}

We made the solid body like spoon by SPH particles. The detailed procedure is
described in the appendix \ref{sec:setup_spoon}.  The total number of the
particle consisting of the spoon was $8652$.  In figure \ref{fig:Mixing:Spoon},
we show the shape of the spoon.  The physical quantities and relative positions
of spoon particles were kept unchanged. The spoon rotates anti-clockwise around
the rotation center of ($0.5,0.5$).  The angular velocity is $2\pi/5$.  We
introduced a repulsive force for the interactions between fluid particles and
solid body particles of the spoon so that penetrations of fluid particles to the
spoon is prevented.

\begin{figure*}[htb]
\begin{center}
\epsscale{1.0}
\plotone{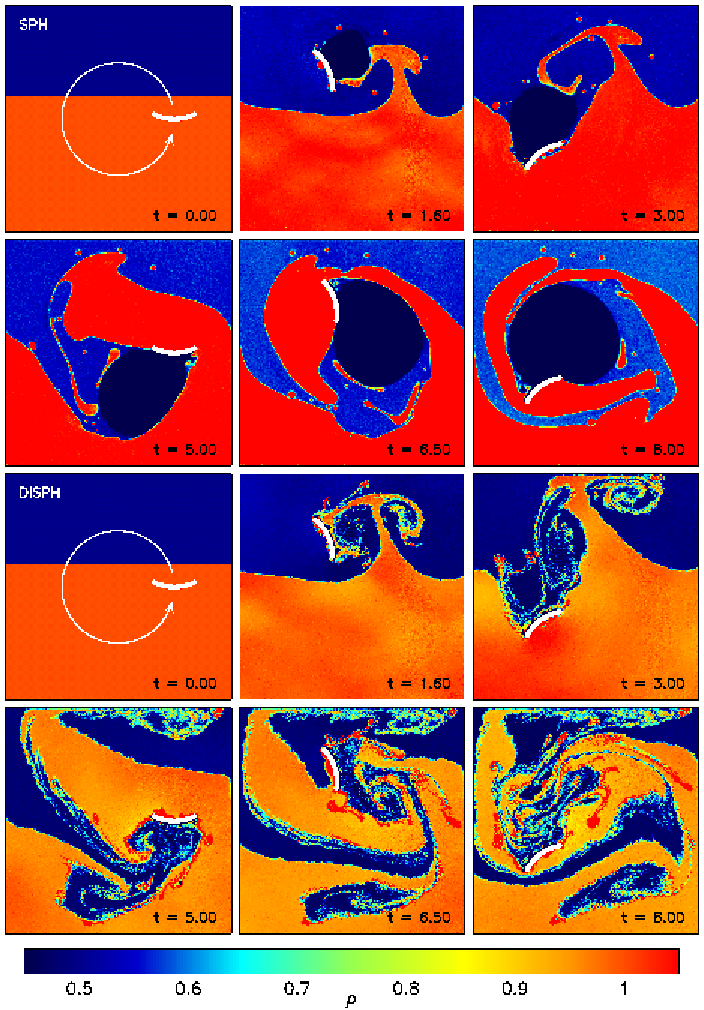}
\caption{Evolution of the two phase fluid forcibly mixed by the spoon up to $t
= 8.0$.  Top six panels show the evolution using the standard SPH, while the
bottom six panels show that using our SPH.  The arcuate structure expressed by
white points is the spoon.  The white long arc (almost circle) found in the
panels at $t = 0$ represents the motion of the spoon and the arrow shows the
direction of the motion (the anti-clockwise).  Colors predict the density of the
fluid and the color bar can be found in the bottom of panels.
\label{fig:Mixing:Snap}
}
\end{center}
\end{figure*}

Figure \ref{fig:Mixing:Snap} shows the snapshots of the representative epoch ($t
= 0, 1.6, 3.0, 5.0, 6.5$, and $8.0$) for the runs with the standard SPH and our
SPH. For the run with the standard SPH, the fluid shows the clear sign of
unphysical surface tension, resulting in the behavior much like that of water
and oil. Although the spoon rotated the two phase fluid, the mixing of two
fluids is prevented by the surface tension.  Overall, the result is completely
different from that obtained by the moving mesh code \citep[see figure 39
in][]{Springel2010}.

The situation was drastically improved when we used DISPH.  When the spoon
lifted-off the dense fluid, a hammer head like structure was developed ($t =
1.6$).  The fluid spilled away from the edges of the spoon and formed eddies.
These prominent structures were not observed in the run with the standard SPH,
while they were observed in the moving mesh simulation \citep{Springel2010}.

\section{Summary and Discussion} \label{sec:summary}

In this paper, we described an alternative formulation of SPH in which the
energy density (pressure), and its arbitrary function  is used as the volume
element instead of the mass density.  In our formulation, the mass of particles
is not used in the evaluation of the right-hand sides of the energy equation and
the equation of motion. As a result, the large error of force estimate at the
contact discontinuity, which is unavoidable with the standard SPH, disappears
completely in our SPH.  Our new SPH includes the Ritchie \& Thomas formulation
\citep{RitchieThomas2001} as a special case.  Not surprisingly, our SPH can
handle contact discontinuities and the Kelvin-Helmholtz and Rayleigh-Taylor
instabilities without difficulty.  The behavior of the shock in DISPH is
essentially the same as that in the standard SPH.  Since the equations used in
our SPH are almost identical to those of in the standard SPH except that a
function of energy density (pressure) is used in place of mass density $\rho$.
Modification of existing SPH code to use our scheme is simple and
straightforward. In particular, there is no increase in the calculation cost at
least for the case of $G(P) = P$.  Equations which are not derived in this
paper, such as the diffusion equation \citep{Brookshaw1985}, can be derived
easily.

\citet{Price2008} improved the treatment of the Kelvin-Helmholtz instability of
the standard SPH, by applying artificial conductivity at the contact
discontinuity. Unlike the artificial viscosity, artificial conductivity
introduces the dissipation not in the original set of equation. Our SPH does not
need such additional dissipation, and thus the contact discontinuity is kept
sharp.

One might think our result contradicts with the requirement that all quantities
in SPH must be smooth \citep{Monaghan1997}. However, it is obvious that in our
SPH, all quantities in the right-hand side of the equations are smooth.  Thus,
our results does not contradict with Monaghan's requirement.

In this paper,we discuss the treatment of ideal gas only. We are currently
working on the extension to non-ideal fluid, and the result will be given in the
forthcoming paper.

\acknowledgements

We thank the anonymous reviewer who provided insightful comments.  We also
thank Evghenii Gaburov, Justin Read, and Takashi Okamoto for helpful comments.
Some of the numerical tests were carried out on the Cray XT4 system in the
Center for Computational Astrophysics at the National Astronomical Observatory
of Japan.  This work is supported by HPCI Strategic Program Field 5 `The origin
of matter and the universe' and Grant-in-Aid for Scientific Research (21244020)
of Japan Society for the Promotion of Science, Ministry of Education, Culture,
Sports, Science and Technology, Japan.

\begin{appendix}

\section{Setup of the Solid Body Like Spoon} \label{sec:setup_spoon}

Here, we describe the procedure to make the spoon which was used in \S
\ref{sec:mixing}.  To express the spoon in the fluid simulation, we first chose
particles in a region of which shape is an arcuate with smooth edges.
Particles in the region evolves like a solid body. We enforced that relative
positions and other physical quantities of particles consisting of the spoon
were kept unchanged during the simulations. To avoid the penetration of fluid
particles to the spoon, we added an repulsive force which acted on the fluid
particles. In below, we explain these in detail.

\begin{figure}[htb]
\begin{center}
\epsscale{1.0}
\plotone{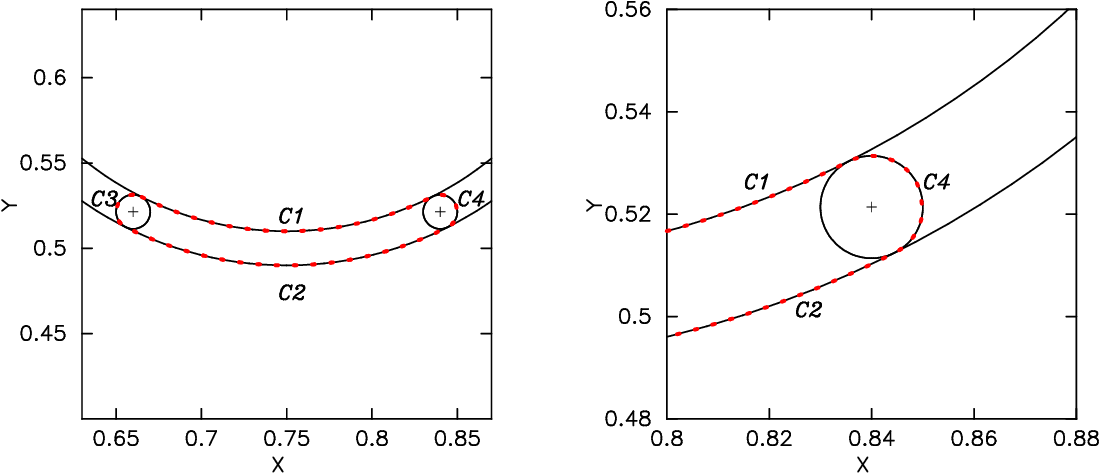}
\caption{Shape of the spoon. Dotted curve with red color exhibits the boundary
of the spoon. Circles consisting of the closed surface are also shown.  {\it
Left}: Whole region of the spoon.  {\it Right}: Close up of the left edge of the
spoon. 
\label{fig:Mixing:Spoon:Shape}
}
\end{center}
\end{figure}

As shown in figure \ref{fig:Mixing:Spoon:Shape}, we combined arcs of four
circles to determine the boundary of the spoon.  We picked up a ringed
region which is located between two circles ({\it C1} and {\it C2}):
\begin{align}
(x-0.75)^2 + (y-0.7)^2 &= (0.19)^2, \\
(x-0.75)^2 + (y-0.7)^2 &= (0.21)^2.
\end{align}
Then, we cut the ringed region and made edges by putting two small circles, {\it
C3} and {\it C4}, which contact to the circles {\it C1} and {\it C2}.  The
centers of two small circles are $(x_{\rm l},y_{\rm l})=(0.66,0.52)$ and
$(x_{\rm r},y_{\rm r})=(0.84,0.52)$, respectively. Their radii are $0.01$. The
right (left) side of {\it C3} ({\it C4}) makes the smooth edge.

The particle found in the enclosed region was selected and converted them into
four smaller SPH particles conserving the center of mass with the separation of
one-half of the original particle separation. The total number of the spoon particle
was $6608$.  The spoon rotated the anti-clockwise around
the rotation center of ($0.5,0.5$) and the angular velocity of $2\pi/5$.

The spoon faced strong pressure from the fluid particles.  When we only
considered the hydrodynamical force from the particles consisting the spoon, a
little but non-negligible amount of fluid particles penetrated through the
spoon.  We, thus, introduced a repulsive force for the interactions between
fluid particles and spoon particles.  In \citet{Monaghan1994},
the Lenard-Jones potential was used for interactions between boundary particles
and fluid particles. Here we choice more simple one.  The form of the repulsive
force between particle $i$ and $j$, $\boldsymbol {F}_{ij}^{\rm rep}$, is the same as the
gravitational force with the Plummer potential:
\begin{equation}
\boldsymbol F_{ij}^{\rm rep} = 
C \frac{m_i m_j}{|r_{ij}^2+\epsilon_{\rm soft}^2|^{3/2}} \boldsymbol r_{ij},
\end{equation}
where $\epsilon_{\rm soft}$ is the softening length of the repulsive force and
is set to $2 h$ of the spoon particles.  We chose $C = 100$.  This force worked
only when fluid particles are in the kernel size of the spoon particles. 
\end{appendix}


\end{document}